\documentclass[10pt,amsmath,amssymb,nofootinbib,twoside,twocolumn,superscriptaddress,floats,floatfix,aps,prd,preprintnumbers]{revtex4-2}

\usepackage{graphicx,epsfig}
\usepackage{amssymb,amsmath,amsthm,amsfonts}
\usepackage{mathrsfs}
\usepackage{bm}
\usepackage[inline]{enumitem}
\usepackage{tensor}
\usepackage[linktocpage]{hyperref}
\usepackage[capitalise]{cleveref}
\usepackage[usenames,dvipsnames]{xcolor}
\usepackage{url}
\usepackage[inline]{enumitem}
\usepackage{xspace}
\usepackage{comment}
\usepackage{ dsfont }
\usepackage[normalem]{ulem}
\usepackage{subfigure}
\hypersetup{
    colorlinks=true,
    linkcolor=red,
    citecolor=blue,
    filecolor=black,
    urlcolor=blue,
    pdfauthor={P. Pani, and A. P. Sanna},
    pdftitle={}
}

\newcommand{\dd}{\text{d}}                              

\newcommand{\ee}{\text{e}}
\newcommand{\ii}{\text{i}}
\newcommand{\OGR}{\omega_\text{GR}}

\newcommand{\sapienza}{Dipartimento di Fisica, Sapienza Università 
	di Roma, Piazzale Aldo Moro 5, 00185, Roma, Italy}
\newcommand{\infn}{INFN, Sezione di Roma, Piazzale Aldo Moro 2, 00185, Roma, Italy}

\allowdisplaybreaks

\begin{document}

\author{Paolo~Pani}
\email{paolo.pani@uniroma1.it}
\affiliation{\sapienza}\affiliation{\infn}

\author{Andrea~P.~Sanna}
\email{asanna@roma1.infn.it}
\affiliation{\infn}

\title{
Scalar shortcut to beyond-Kerr ringdown tests\\ and their complementarity with black-hole shadow observations}

\begin{abstract}
The quasinormal modes of black holes~(BHs) in the large-angular-momentum limit can be computed within the eikonal approximation. This approximation is often extrapolated to low angular momentum to obtain a rough estimate of the dominant ringdown modes. Although approximate, this approach is particularly convenient in theories beyond general relativity with intricate dynamics, or for phenomenological metrics that lack an underlying fundamental theory.
Here we explore a complementary approximate strategy: we compute exactly the quasinormal modes of a test scalar field propagating on the BH background and use their \emph{deviations} from the general-relativity predictions as a proxy for the corresponding corrections to the gravitational quasinormal modes.
For Kerr–Newman and Einstein-scalar–Gauss–Bonnet BHs, we show that this method reproduces the exact corrections (including the coupling among different degrees of freedom) within
tens of percent, an accuracy that is adequate as long as ringdown measurements remain at the percent level.
Furthermore, this method is typically comparable to, or more accurate than, the eikonal approximation.
We then apply the same strategy to phenomenological metrics commonly employed in tests of gravity using BH imaging. By computing scalar quasinormal modes in a large family of these metrics for the first time, we find that current ringdown constraints are comparable to, and in some cases more stringent than, those derived from BH shadow observations, while also providing complementary bounds on sectors that would otherwise be inaccessible.
\end{abstract}

\maketitle

\section{Introduction}
\label{sec:Intro}

The detection of gravitational waves~(GWs) from compact binary mergers by the LIGO–Virgo–KAGRA collaboration has opened an unprecedented window onto the strong-field regime of gravity~\cite{LIGOScientific:2016aoc,LIGOScientific:2016sjg,LIGOScientific:2020ibl}. 
In particular, the \emph{ringdown} phase of a binary black-hole~(BH) coalescence, namely the stage during which the remnant relaxes to equilibrium through damped oscillations, provides a clean probe of the geometry in the immediate vicinity of the horizon. 
Within general relativity~(GR), the ringdown signal is described by a superposition of quasinormal modes~(QNMs) whose frequencies and damping times depend solely on the mass and spin of the remnant Kerr BH~\cite{Vishveshwara:1970zz,Press:1971wr,Teukolsky:1973ha,Chandrasekhar:1975zza,Berti:2009kk}. 
This property is a consequence of the uniqueness of the Kerr solution and enables stringent null tests of GR and of the nature of compact objects through BH spectroscopy~\cite{Dreyer:2003bv,Berti:2005ys,Gossan:2011ha,Berti:2016lat,Cardoso:2019rvt,Maggio:2021ans,LIGOScientific:2025rid,LIGOScientific:2025wao} (see~\cite{Berti:2025hly} for a recent overview).

The increasing sensitivity of current detectors and the prospects offered by next-generation observatories such as the Einstein Telescope~\cite{ET:2019dnz,Branchesi:2023mws,ET:2025xjr}, Cosmic Explorer~\cite{Reitze:2019iox,Evans:2021gyd}, and space interferometers such as LISA~\cite{LISA:2024hlh} motivate the development of accurate theoretical frameworks to parameterize and constrain possible deviations from the Kerr metric in the ringdown regime. 

A key challenge in this context is that QNMs are global properties of the spacetime, determined by boundary conditions at the horizon and at infinity and by the actual spacetime geometry of the remnant. 
Therefore, even small modifications of the geometry can lead to observable shifts in the QNM spectrum, potentially encoding signatures of new physics. While this provides a unique portal to strong gravity and near-horizon physics, computing QNMs beyond the standard Kerr paradigm in GR is challenging.
The QNM corrections generically depend on the remnant mass and spin, as well as on any fundamental coupling constant characterizing a given extension of GR or on deformation parameters describing departures from the Kerr metric. In a modified theory of gravity, their explicit form must be computed on a case-by-case basis, either perturbatively in the spin~\cite{Pani:2009wy,Cardoso:2009pk,Molina:2010fb,Pani:2013ija,Pani:2013wsa,Pierini:2021jxd,Wagle:2021tam,Cano:2021myl,Pierini:2022eim,Cano:2023jbk,Wagle:2023fwl,Cano:2024jkd,Cano:2024ezp} or fully numerically~\cite{Dias:2015wqa,Chung:2023zdq,Chung:2023wkd,Blazquez-Salcedo:2023hwg,Chung:2024ira,Chung:2024vaf}.
Both computations are quite demanding and must be performed for any desired theory.\footnote{An alternative strategy is to parametrize deviations in the QNM spectrum in a phenomenological way~\cite{LIGOScientific:2021sio,Ma:2023cwe,Ma:2022wpv}, possibly guided by explicit examples of solutions in theories beyond GR~\cite{Maselli:2019mjd}. 
However, this approach typically introduces a large number of free parameters. Moreover, interpreting the resulting bounds ultimately requires input from specific theories in order to relate the phenomenological coefficients to the underlying fundamental parameters.}

Another widely adopted strategy to search for deviations from GR is to consider \emph{phenomenological metrics} that parametrize generic deformations of the Kerr solution while preserving key symmetries such as stationarity and axisymmetry~\cite{Johannsen:2011dh,Johannsen:2013rqa,Johannsen:2013szh,Cardoso:2014rha,Konoplya:2016jvv}. 
Such metrics do not necessarily arise as consistent solutions to a given modified theory of gravity (although they may be interpreted as GR solutions in the presence of exotic matter fields). Their main scope is for tests of gravity which do not involve the dynamical regime, e.g. X-ray reflection spectroscopy and BH shadow observations~\cite{Bambi:2015kza,EventHorizonTelescope:2019dse,EventHorizonTelescope:2022wkp}. 
The latter, in particular, have provided horizon-scale images of the supermassive BHs in M87 and Sgr~A*, enabling constraints on deviations from Kerr based on the size and shape of the photon ring~\cite{EventHorizonTelescope:2019ggy,EventHorizonTelescope:2022wkp}. 
These measurements probe null geodesics in the near-horizon region and are therefore sensitive to the light-ring structure of the spacetime~\cite{Cardoso:2019rvt}, but not to its dynamics.

Ringdown GWs offer a complementary probe. 
In the eikonal (namely, large-angular-momentum) limit, QNM frequencies are closely related\footnote{See also~\cite{Konoplya:2017wot,Konoplya:2022gjp} for examples where light propagation does not follow the spacetime metric and for discussions on the limitations of the eikonal approximation in capturing the full QNM spectrum}. to the properties of unstable photon orbits~\cite{Ferrari:1984zz,Cardoso:2008bp,Stefanov:2010xz}. Owing to its analytical simplicity, this correspondence has already been employed as a practical tool for ringdown tests of GR using parametrically-deformed spacetimes~\cite{Volkel:2020daa,Dey:2022pmv}.
Beyond this approximation, however, the QNM spectrum depends on the full structure of the perturbation potential and on possible couplings among different perturbative degrees of freedom.
Consequently, ringdown spectroscopy can access aspects of the geometry that are not directly constrained by shadow measurements, including the structure of the effective potential governing gravitational perturbations. 
Establishing a consistent framework to compare constraints from ringdown and from BH shadows is therefore of primary importance. 

In this work, we develop a method to compute the ringdown spectrum of generic stationary BH spacetimes described by either alternative gravity theories with extra degrees of freedom or by phenomenological metrics. 
The idea is to compute exactly the QNMs of a \emph{test} field propagating on a given BH background and use their \emph{deviations} from the GR predictions as a proxy for the corresponding corrections to the more involved gravitational QNMs.
We validate this method against numerical results for specific models, showing that even the simplest test scalar-field model reproduces the exact QNM shifts within tens of percent accuracy, even in scenarios where mode mixing is present in the full set of equations. 
Such precision is adequate for current and near-future ringdown measurements, whose statistical uncertainties remain at the percent level.

Both the eikonal and test-scalar field approaches share the advantage that the corresponding QNM frequencies can be computed once the background metric is specified, without requiring knowledge of the full gravitational field equations. They should therefore be regarded as geometric probes of the background spacetime, in contrast with gravitational QNMs, whose spectrum depends on the dynamics of the underlying theory through the perturbation equations. The test-scalar approach is computationally more demanding than the eikonal approximation, since it requires solving the full wave equation with QNM boundary conditions. Moreover, unlike in GR, beyond-GR spacetimes rarely allow the radial-angular separation of the perturbation equations, substantially complicating the calculation compared with extracting the properties of the unstable light ring. However, at variance with the eikonal approximation, it is nonlocal, as it relies on the full spacetime geometry and on the global boundary conditions defining the QNM problem. Therefore, it also retains information about the global structure of the effective potential and becomes significantly more accurate whenever the eikonal approximation is not quantitatively reliable. Finally, it shares with the eikonal approximation the property of becoming exact in the large-angular-momentum limit, since in this regime the eikonal QNMs of a test scalar field coincide with the exact ones.

Our approach is therefore sufficiently general to accommodate parametrized deviations from Kerr, while remaining accurate enough to reproduce the exact results in known cases. It also enables a direct mapping between deviations in the background metric and shifts in observable QNM frequencies. 
This mapping can be used to place bounds on the parameters of phenomenological metrics using ringdown observations, in close analogy with existing shadow-based analyses. 
By comparing the parametric dependence of QNM frequencies and photon-ring observables on the same deformation parameters, one can assess the complementarity and potential degeneracies between GW and electromagnetic probes. 
In particular, while shadow measurements primarily constrain the properties of null geodesics near the light ring, ringdown modes are sensitive to the full structure of the perturbation equations, including near-horizon boundary conditions. 
A combined analysis therefore offers a more robust and multidimensional test of the Kerr hypothesis~\cite{Volkel:2020xlc}.

Our results pave the way toward a unified framework in which constraints from ringdown spectroscopy and from BH shadows can be consistently compared within the same parametrized deviations from Kerr. 
As observational capabilities improve, especially with next-generation GW detectors and next-generation very-long-baseline interferometry, such joint analyses will play a crucial role in testing the nature of astrophysical BHs and in probing possible departures from GR in the strong-field regime.

The rest of the paper is organized as follows. In \cref{Sec:TradingQNMdeviations}, we investigate whether deviations from the Kerr metric, encoded in test scalar-field modes, can serve as proxies for the gravitational predictions. We then examine two well-motivated modified Kerr geometries. In \cref{Sec:KerrNewman_ScalarvsGravModes}, we study the Kerr–Newman spacetime and perform a detailed comparison, including the eikonal results. \Cref{Sec:EdGB_ScalarvsGravModes} contains a more focused comparison between test scalar and gravitational results in Einstein-scalar-Gauss-Bonnet gravity. In \cref{Sec:JohannsenMetric}, we apply our test-scalar approach to compute semi-analytically the exact scalar modes of the Johannsen metric, a phenomenologically modified Kerr spacetime widely used in BH imaging tests. After reviewing the metric and current observational constraints, we outline the algorithm used to solve the massless Klein–Gordon equation and compute scalar QNMs for cases where individual deformation parameters are activated. We also discuss the complementarity with existing Event Horizon Telescope~(EHT) bounds and interpret the results using the eikonal approximation. We finally draw our conclusions in \cref{sec:Conclusions}. Additional technical discussions and results are presented in four appendices. 

Throughout the paper, we adopt units in which $c = G = 1$. We will label the QNMs as $(n, \ell, m)$, with $n$, $\ell$ and $m$ the overtone, polar and azimuthal numbers, respectively. We will also focus on $\ell = |m|$ modes which are often the dominant ones in the ringdown.

\section{Trading QNM deviations}
\label{Sec:TradingQNMdeviations}

The goal of this section is to assess to what extent the impact of deviations from a reference spacetime on gravitational QNMs can be reliably estimated using either a test scalar-field\footnote{We focus on the simplest case of a test \emph{scalar} perturbation, but the method can be applied also to test spin-1 and, especially, spin-2 perturbations. While more complicated to deal with, we expect better agreement with the exact case compared to the scalar-field model.}  model or the eikonal approximation.

\subsection{Tolerance on approximated GR deviations}
It is important to stress that we do not aim to replace the gravitational QNMs entirely with their scalar or eikonal counterparts. Rather, our strategy is to trade only the \emph{deviations} from the GR predictions in the gravitational sector with the corresponding deviations computed in the scalar sector or within the eikonal approximation.

Throughout most of this work, we adopt the Kerr spacetime as the reference background and denote its QNM frequencies by $\OGR$. For our purposes, $\OGR$ is treated as an exactly known reference quantity, assuming the background parameters to be known without uncertainty. This differs from realistic data-analysis scenarios, where a single measurement may encode degeneracies between beyond-GR effects and the underlying background parameters. The resulting correlations will not be considered here, as our focus is on the relative deviations from the GR predictions and on how they are modeled by the various modes, rather than performing parameter estimation or using real data. The frequencies of the modified geometries are, instead, denoted by $\omega$. To quantify how accurately test scalar or eikonal modes capture the deviations predicted by gravitational perturbations, we parametrize the modified frequencies as 
 \begin{equation}
     \omega \equiv \OGR(1 + \Delta \omega)\, ,
     \label{eq:RelativeDiscr}
 \end{equation}
where $\Delta \omega$ represents the relative deviation with respect to the Kerr results. This parametrization isolates the genuine response of the QNM spectrum to deformations of the background spacetime and allows for a direct comparison between different perturbative sectors.

The use of eikonal or test scalar-mode deviations as proxies for gravitational ones is meaningful only insofar as the resulting discrepancies between the different predictions remain smaller than or comparable with the accuracy currently achievable by ringdown measurements. If the mismatch between scalar/eikonal and gravitational results is smaller than present observational uncertainties, then the former provide an operationally reliable proxy for the latter. To formalize this criterion, we introduce an observationally motivated tolerance band around the gravitational results for $\Delta \omega$. The width of this band can be estimated by computing the absolute error $\delta \Delta \omega$ on the relative deviation $\Delta \omega$ defined in \cref{eq:RelativeDiscr}. Additionally, throughout this section, $\delta \omega$ and $\delta \OGR$ will denote absolute uncertainties on the modified and Kerr QNM frequencies, respectively. 
From \cref{eq:RelativeDiscr}, combining relative errors in quadrature and neglecting correlations for simplicity, we get
\begin{equation}
    \frac{\delta \omega}{\omega} = \sqrt{\left(\frac{\delta \OGR}{\OGR}\right)^2+\left(\frac{\delta \Delta\omega}{1+\Delta\omega}\right)^2}\,.
    \label{eq:RelativeErrorsOmegaTotal}
\end{equation}
The inclusion of correlations between the Kerr frequency and the inferred deviation would introduce in \cref{eq:RelativeErrorsOmegaTotal} a covariance term which could broaden or narrow the resulting tolerance band, depending on its sign.
Although this effect could quantitatively modify the results reported here, we do not expect it to affect our qualitative comparison among scalar, eikonal, and gravitational deviations.

Since in our analysis the reference Kerr spectrum is treated as exactly known, we set $\delta\omega_{\rm GR}=0$. Equation~\eqref{eq:RelativeErrorsOmegaTotal} then allows us to translate a measurement of a mode with relative accuracy $\delta\omega/\omega<X$ into an upper bound on the \emph{absolute} uncertainty of the deviation,
\begin{equation}
    \delta \Delta\omega < X\,|1+\Delta \omega|\, ,
    \label{eq:BandWidth}
\end{equation}
which in turn translates into an upper bound on its \emph{relative} uncertainty
\begin{equation}
    \frac{\delta \Delta\omega}{|\Delta\omega|} < X\frac{|1+\Delta \omega|}{|\Delta\omega|}\, .
\end{equation}
This condition reflects a simple, but important aspect: smaller deviations $\Delta \omega$ require less stringent accuracy, since their impact on the total uncertainty on $\omega$ is  correspondingly suppressed. 

The tolerance band defined by $X$, therefore, provides a practical criterion to judge whether scalar or eikonal modes can serve as faithful estimators of the gravitational spectrum. As a reference value for $X$, we consider current constraints on deviations from the Kerr metric inferred from GW250114, the loudest GW event observed to date~\cite{LIGOScientific:2025rid}. In particular, this event constrains deviations in the $(n,\ell,m)=(0, 2, 2)$ mode at the level of approximately $4\%$ for the real part and $10\%$ for the imaginary part of the frequency~\cite{LIGOScientific:2025wao}. Despite the different sensitivities to these two components, we conservatively select the tighter bound and adopt $X = 4\%$, for both real and imaginary parts throughout the following sections.

As representative examples of modified Kerr geometries, we consider the Kerr-Newman family in \cref{Sec:KerrNewman_ScalarvsGravModes} and rotating solutions in shift-symmetric Einstein-scalar-Gauss-Bonnet (EsGB) gravity in \cref{Sec:EdGB_ScalarvsGravModes}. In each case, we compare the deviations $\Delta \omega$ predicted by the eikonal and scalar modes with those obtained from gravitational perturbations, thereby assessing whether these simpler approximations reproduce the gravitational corrections with an accuracy compatible with current observational sensitivities.
\subsection{Kerr-Newman spacetime}
\label{Sec:KerrNewman_ScalarvsGravModes}
The Kerr-Newman line element reads~\cite{Newman:1965my}
\begin{equation}
\begin{split}
    \dd s^2 = &-\left[1-\frac{2M r-Q^2}{\Sigma(r, \theta)} \right]\dd t^2 + \frac{\Sigma(r, \theta)}{\Delta(r)} \dd r^2 \\
    &+ \Sigma(r, \theta) \dd \theta^2 - \frac{4Mr-2Q^2}{\Sigma(r, \theta)}\,  \dd t \, \dd \phi  \\
    &+ \left[(r^2 + a^2)\sin^2\theta + \frac{2Mr-Q^2}{\Sigma(r, \theta)} a^2 \sin^4 \theta \right]\dd \phi^2\, ,
\end{split}
\end{equation}
where the metric functions are
\begin{equation}
\begin{split}
    \Sigma(r, \theta) &= r^2 + a^2 \cos^2\theta \, ; \\
    \Delta(r) &= r^2 - 2M r+a^2 + Q^2\, .
\end{split}
\end{equation}
Here, $M$ and $Q$ denote the mass and the electric charge of the BH, while $a$ is the spin parameter. The Kerr-Newman metric is the most general, stationary, axisymmetric and asymptotically-flat solution of Einstein-Maxwell theory describing a rotating, charged BH. As such, it provides a theoretically well-controlled setting in which deviations from the Kerr geometry can be investigated while remaining entirely within the framework of standard GR. In practice, we treat $Q$ as a deformation of the Kerr background induced by the presence of an additional external field (the electric one). We explore values of the charge-to-mass ratio $Q/M$ in the range $[0,0.8]$, while fixing the dimensionless spin to the reference value $a/M = 0.5$. Additional results for $a/M = 0$ and $a/M = 0.2$ are presented in Appendix~\ref{App:KNComparisonAddition}.

Our analysis focuses on the three modes $\{(0,2,2),\,(1,2,2),\,(0,3,3)\}$, which are the longest-lived ones and, therefore, most relevant during the ringdown phase and for BH spectroscopy~\cite{JimenezForteza:2020cve}. Using \cref{eq:RelativeDiscr}, we compare the deviations $\Delta \omega$ encoded in the gravitational modes with those predicted by test-scalar perturbations and by the eikonal approximation. Owing to its analytical simplicity, the latter is often adopted in analyses of phenomenologically-modified BHs and provides a useful benchmark against which approximate descriptions of the QNM spectrum can be  assessed~\cite{Press:1971wr,Ferrari:1984zz,Mashhoon:1985cya,Kokkotas:1999bd,Cardoso:2008bp,Berti:2009kk,Stefanov:2010xz,Yang:2012he,Konoplya:2017wot,Glampedakis:2017dvb,Glampedakis:2019dqh,Silva:2019scu,Bryant:2021xdh}. 

For the test-scalar QNMs, we use the data from Ref.~\cite{Berti:2005eb} for the $(0, 2, 2)$ and $(1, 2, 2)$ modes. The $(0, 3, 3)$ scalar frequencies are, instead, evaluated using the Leaver continued-fraction method~\cite{Leaver:1985ax}. Gravitational QNM frequencies~\cite{Dias:2015wqa} are obtained from the numerical fits presented in Ref.~\cite{Carullo:2021oxn}, which provide analytical representations of the frequencies as functions of $a/M$ and $Q/M$, allowing for a more direct and efficient evaluation. Eikonal QNM frequencies are evaluated using the formulae summarized in the brief review contained in Appendix~\ref{App:EikonalApproximation}.

\Cref{fig:KN_all} shows plots of the absolute value of the relative deviations, computed using \cref{eq:RelativeDiscr}, as functions of $Q/M$.
\begin{figure*}[h!]
    \centering
    \includegraphics[width=0.32\textwidth]{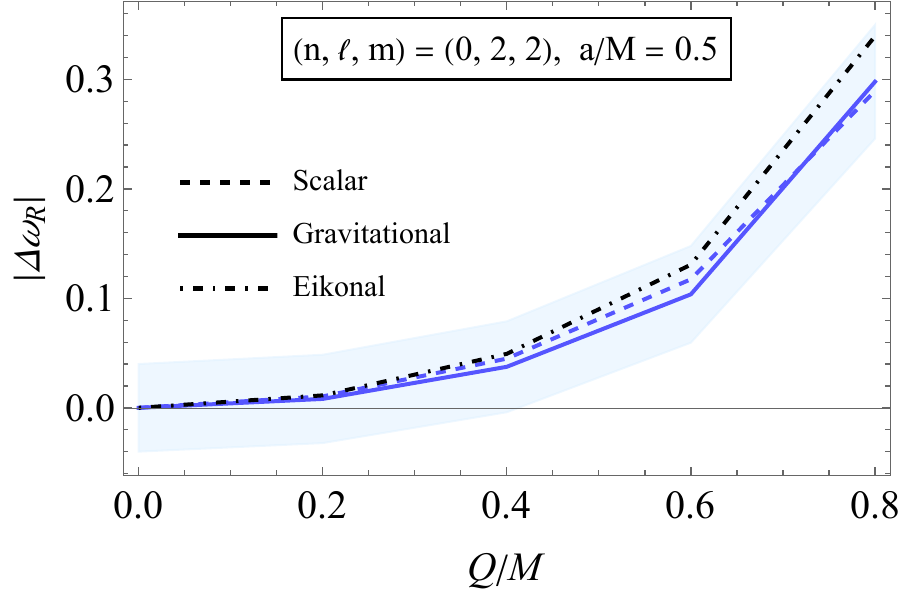}
    \includegraphics[width=0.32\textwidth]{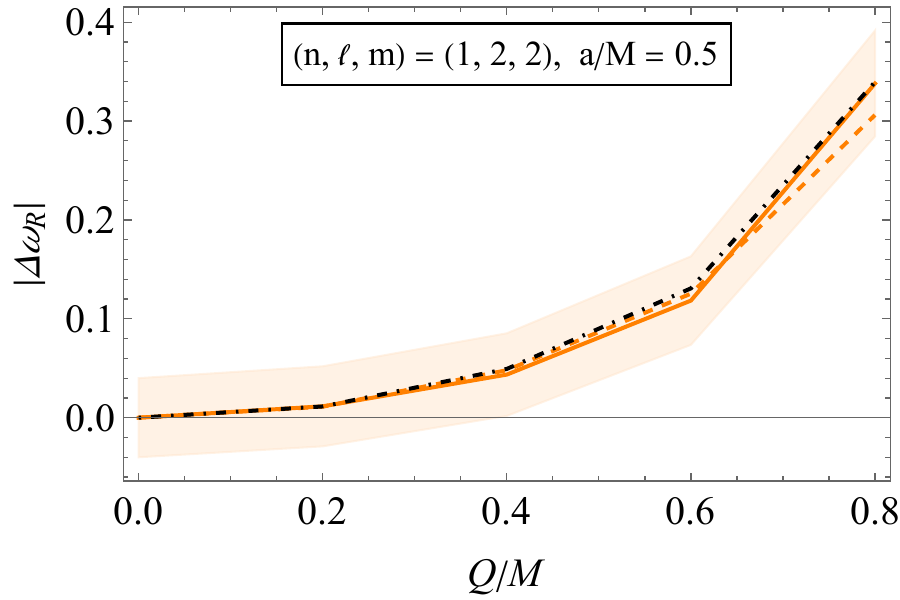}
    \includegraphics[width=0.32\textwidth]{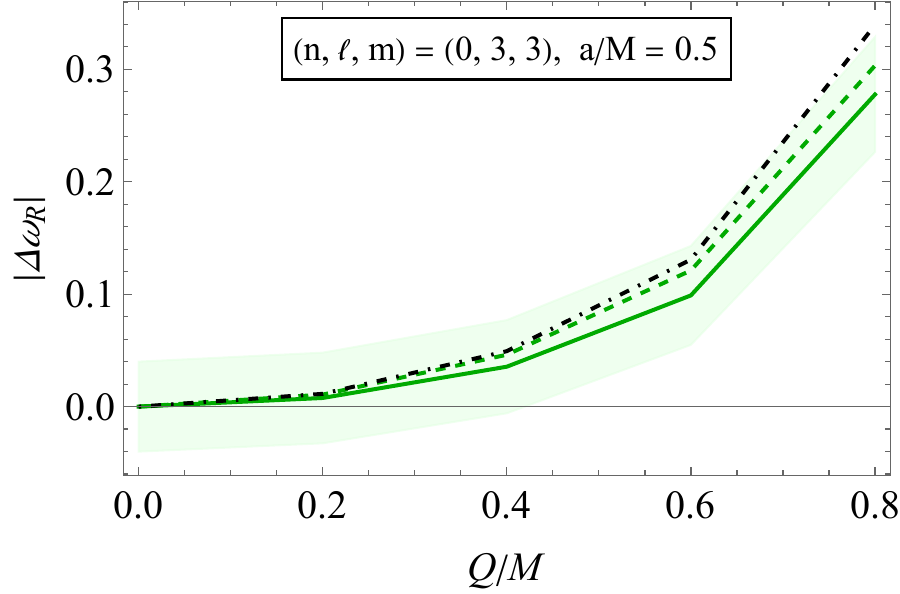}\\
    \includegraphics[width=0.32\textwidth]{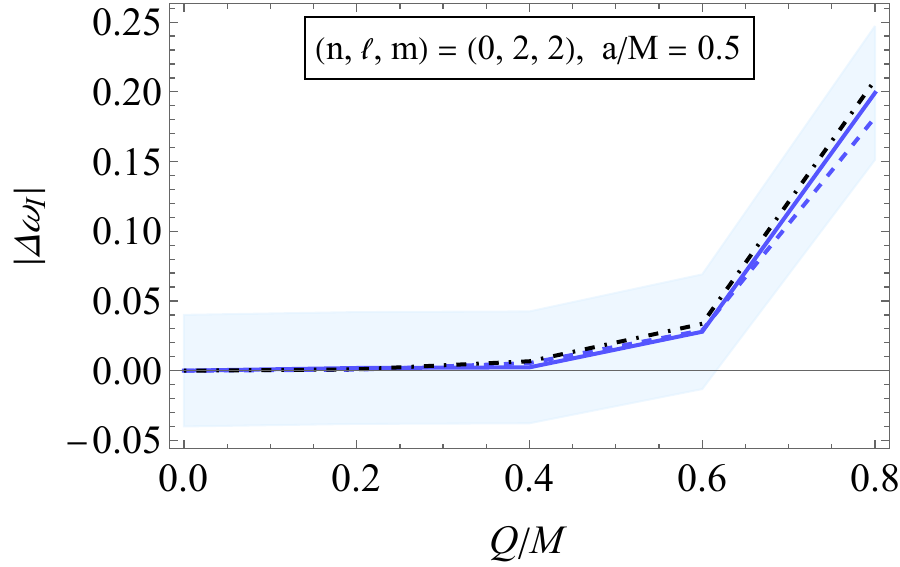}
    \includegraphics[width=0.32\textwidth]{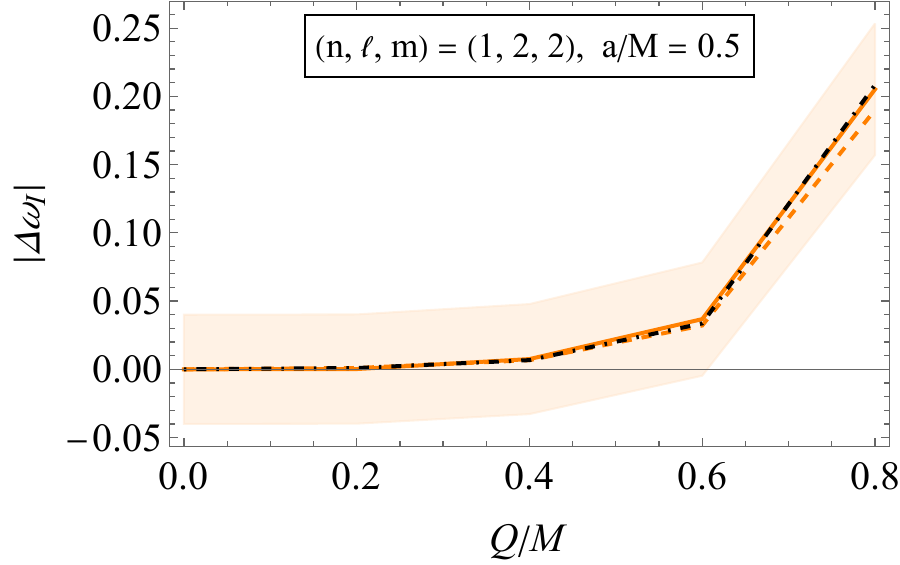}
    \includegraphics[width=0.32\textwidth]{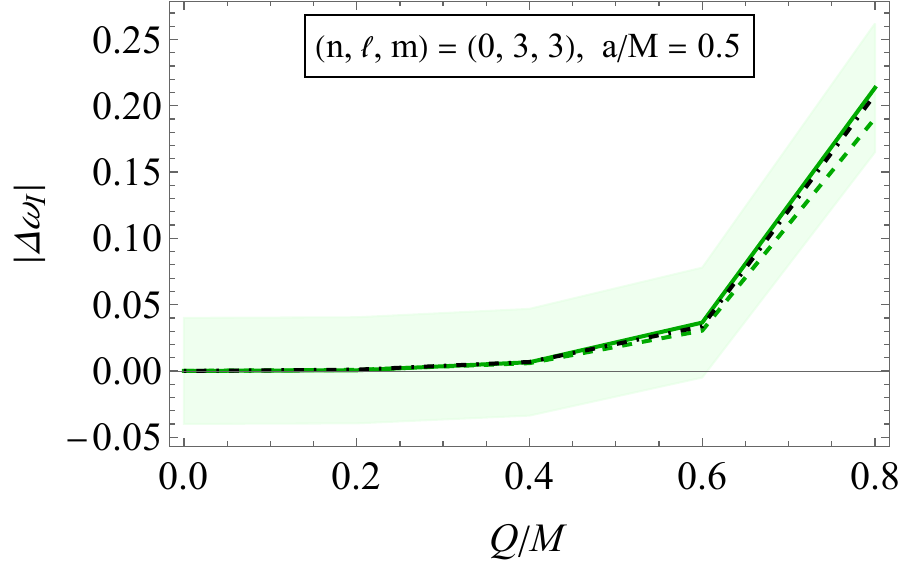}\\
    \caption{Plots of the absolute values of the relative deviations of Kerr--Newman QNMs from their Kerr counterparts as functions of $Q/M$. Both real and imaginary parts are shown. In all panels, the spin is fixed to $a/M = 0.5$. Solid and dashed lines correspond to gravitational and scalar results, respectively, while the dot-dashed black lines denote eikonal predictions. Shaded regions represent observationally motivated tolerance bands around the gravitational results, whose widths are given by the right-hand side of \cref{eq:BandWidth} with $X = 4\%$.}
    \label{fig:KN_all}
\end{figure*}
Shaded bands correspond to observationally motivated tolerances with $X = 4 \, \%$ in \cref{eq:BandWidth}.

We find that, even for moderately large values of $Q/M$, scalar and eikonal predictions of deviations from the Kerr spectrum closely follow the gravitational ones well within the tolerance bands. The only exception occurs for the eikonal predictions of the real part of the $(0, 3, 3)$ mode, which slightly exceed the band for $Q/M \gtrsim 0.6$. 

The agreement is particularly striking, since 
gravitational perturbations of a Kerr-Newman BH are coupled to the electromagnetic ones and are notoriously difficult to solve, as the corresponding set of partial differential equations do not separate in a radial and angular part~\cite{Dias:2015wqa}.

\subsection{Rotating solutions in Einstein-scalar-Gauss-Bonnet}
\label{Sec:EdGB_ScalarvsGravModes}

As a further example of axisymmetric BHs beyond GR, we consider rotating solutions in shift-symmetric EsGB gravity~\cite{Delgado:2020rev,Sullivan:2020zpf}. The Lagrangian density of the theory reads
\begin{equation}
\mathcal{L} = \frac{\sqrt{-g}}{16\pi}\left[\mathcal{R}-\frac{1}{2}(\nabla_\mu \phi)^2 + \alpha \, \phi \mathcal{G}\right]\, ,
\label{eq:GaussBonnetLagrangian}
\end{equation}
where $\mathcal{R}$ denotes the Ricci scalar, $\phi$ is a real scalar field and $\mathcal{G} = \mathcal{R}_{\mu\nu\rho\sigma}\mathcal{R}^{\mu\nu\rho\sigma}-4\mathcal{R}_{\mu\nu}\mathcal{R}^{\mu\nu} + \mathcal{R}^2$ is the Gauss--Bonnet invariant, non-minimally coupled to $\phi$ through the linear interaction term $\alpha \phi$. The coupling constant $\alpha$ has dimensions of length squared. It is convenient to introduce the dimensionless parameter $\xi \equiv \alpha/M^2$, where $M$ is the BH mass, which we adopt as the reference scale. In the limit $\xi = 0$, the Kerr metric is recovered, consistent with no-hair theorems. For nonzero $\xi$, instead, the scalar-curvature coupling induces deviations from the Kerr geometry, which should impact the observed phenomenology, including the GW waveforms. This leads to the possibility of observationally testing the theory. The most stringent bounds on $\xi$, derived from GW observations, require $\sqrt{\alpha} \lesssim \mathcal{O}(1) \, \text{km}$ (see, e.g., Refs.~\cite{Lyu:2022gdr,Wang:2023wgv} and references therein). For typical stellar-mass BHs involved in binary mergers, $M \sim 5-10 \, M_\odot$, this translates into $\xi \lesssim \mathcal{O}(10^{-2})$. 
\begin{figure*}
    \centering
    \includegraphics[width=0.32\textwidth]{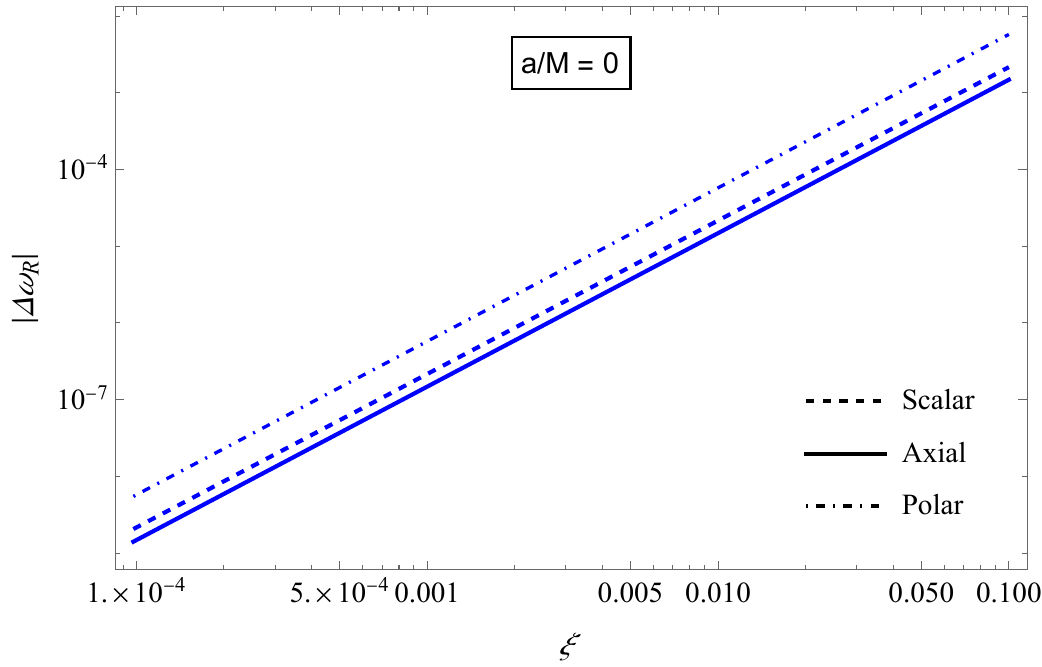}
    \includegraphics[width=0.32\textwidth]{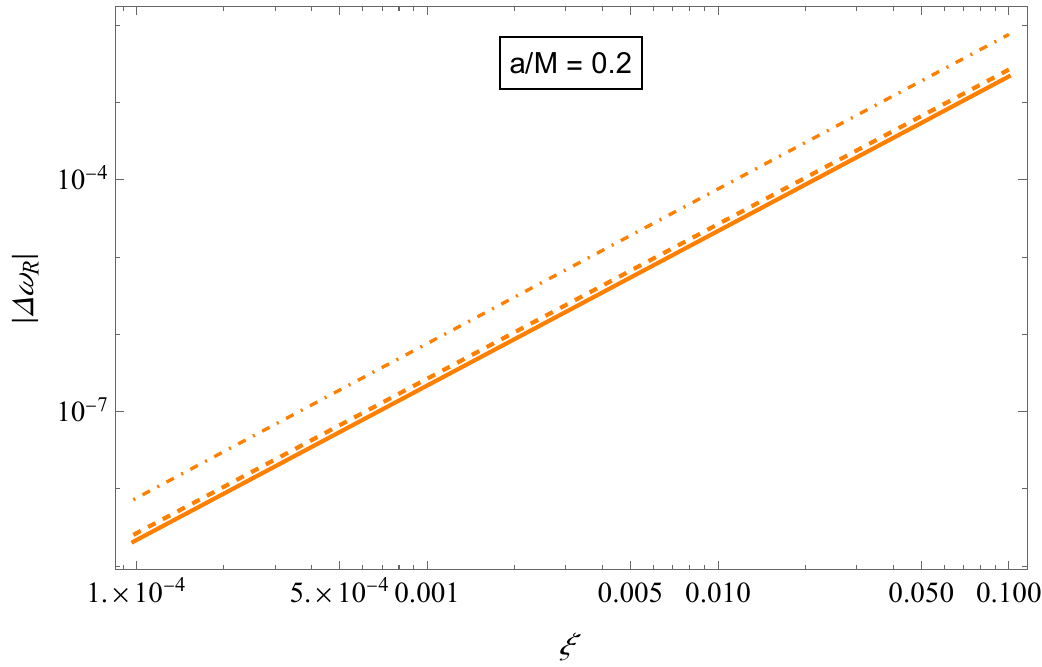}
    \includegraphics[width=0.32\textwidth]{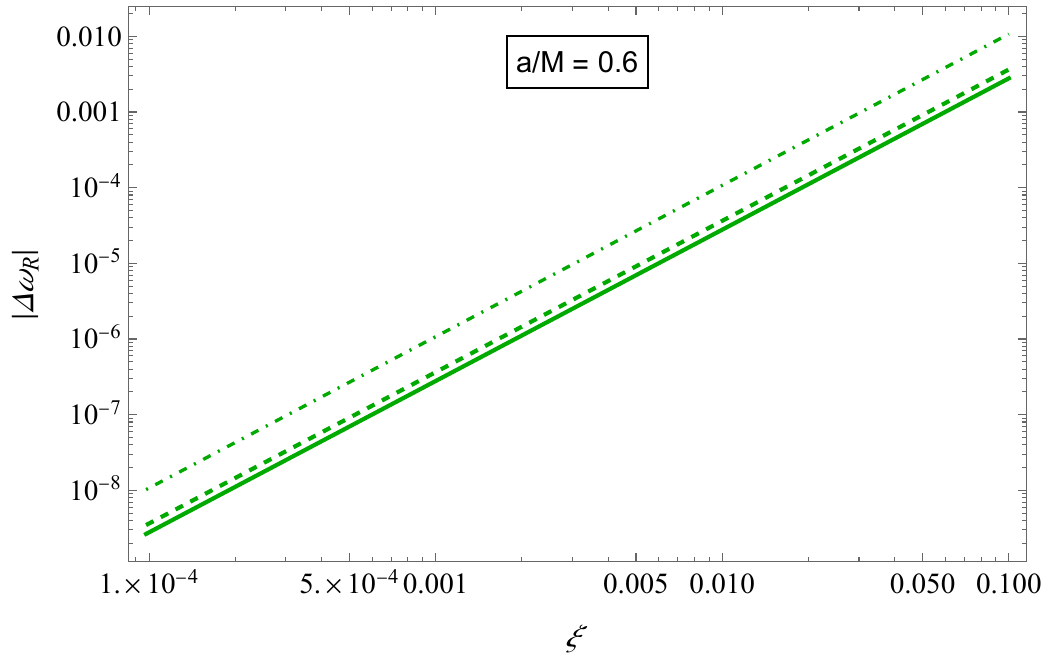}\\
    \includegraphics[width=0.32\textwidth]{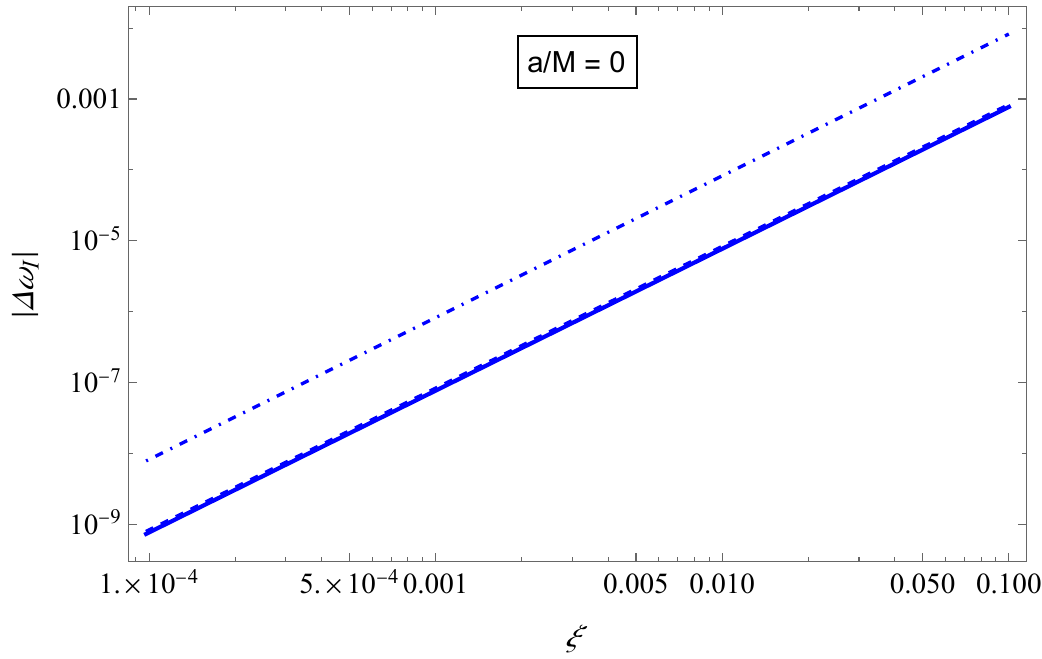}
    \includegraphics[width=0.32\textwidth]{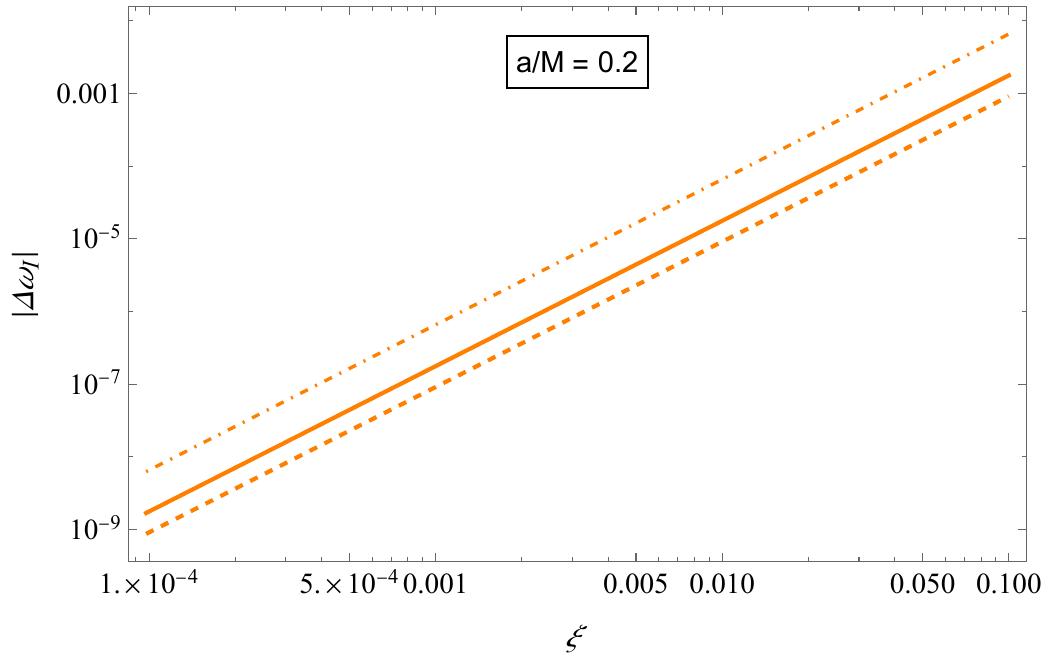}
    \includegraphics[width=0.32\textwidth]{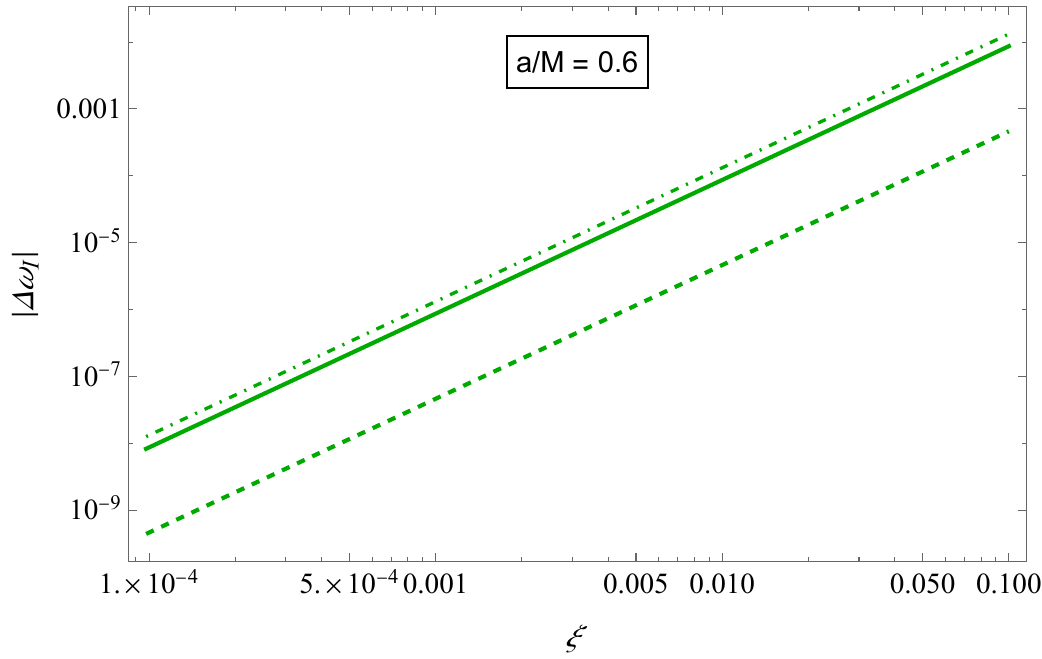}\\
    \caption{Log--log plots of the absolute values of the relative deviations of the frequencies of the mode $(0, 2, 2)$ for rotating BHs in shift-symmetric EsGB gravity from their Kerr counterparts. Both real and imaginary parts are shown as functions of the dimensionless coupling constant $\xi$, for the representative values $a/M = 0, \, 0.2, \, 0.6$. Dashed, solid and dot-dashed lines correspond to test-scalar, axial and polar gravitational results, respectively.}
    \label{fig:EdGB}
\end{figure*}
In what follows, as we are interested in the ringdown phase, we study deviations from the Kerr QNM spectrum as functions of the coupling constant $\xi$. Since the present observational constraints hint at small values of $\xi$, we will focus on the small-coupling perturbative regime of the theory. Consequently, the present discussion should be interpreted as an order-of-magnitude comparison and as an evaluation of the accuracy of our approximation strategy. 
Since earlier analyses did not reveal qualitative differences among different multipoles, and in order to capture the main phenomenological features while keeping the discussion focused, we restrict attention to the fundamental $(0,2,2)$ mode, which currently provides the strongest ringdown constraints~\cite{LIGOScientific:2025wao}.

Again we consider scalar and gravitational modes. Eikonal results are not included here, as the previous analyses revealed that they do not exhibit qualitative differences with respect to test-scalar perturbations. 

In the following, by “scalar modes” we refer to the QNMs of a minimally coupled test scalar field propagating on the EsGB BH background described as a solution to the theory~\eqref{eq:GaussBonnetLagrangian}, and not to the perturbations of the scalar field $\phi$. Indeed, the latter are coupled to the gravitational one, making the exact gravitational QNM computation very challenging~\cite{Khoo:2024agm}. Test scalar field QNMs were computed in Ref.~\cite{Cano:2020cao}, where the most general class of higher-derivative gravity theories up to quartic order in the curvature was considered, with EsGB as a specific subclass. The frequencies of fundamental modes with $\ell = 0,\, 1,\, 2$ are obtained numerically by solving the scalar perturbation equations, using a projection onto spheroidal harmonics to handle the non-separability of the scalar wave operator. From these numerical results, analytic fitting formulas were constructed as functions of the Gauss–Bonnet coupling $\xi$ and the dimensionless spin parameter $a/M$. The fits are calibrated up to $a/M \lesssim 0.7$, beyond which the polynomial expansions gradually lose accuracy.

Fits for the gravitational QNM frequencies of rotating EsGB BHs were instead reported in Ref.~\cite{Chung:2024vaf} for arbitrary spins  (but see Refs.~\cite{Cano:2021myl,Pierini:2021jxd,Pierini:2022eim} for results in the slow-rotation limit).

In the following analysis, we consider three representative values of the dimensionless BH spin: $a/M = 0, \, 0.2, \, 0.6$. A distinctive feature of the gravitational QNM spectrum in EsGB gravity is the breaking of isospectrality between axial and polar modes for $\xi \neq 0$. We thus consider both sectors. We consider values of the coupling up to $\xi \lesssim \mathcal{O}(10^{-1})$.

The results are shown in \cref{fig:EdGB}. The deviations from the Kerr spectrum in both scalar and gravitational sectors display the same qualitative behavior, scaling approximately as $\xi^2$ in the small-coupling regime, as expected by treating the theory as a small deformation away from GR. 
Moreover, axial gravitational modes systematically lie closer to the test-scalar predictions than polar modes. This behavior is also expected, since polar perturbations couple directly to the perturbations of the scalar field $\phi$, whereas axial perturbations remain decoupled. 

Despite these sector-dependent differences, the overall picture is qualitatively similar to that found for the Kerr–Newman spacetime. In particular, we have verified that scalar predictions for the deviations from the Kerr QNM spectrum closely track both axial and polar gravitational results, remaining well within the tolerance bands\footnote{The tolerance bands are not explicitly displayed in \cref{fig:EdGB}, as including them would significantly clutter the plots without adding substantial information. } introduced in the previous section and defined by \cref{eq:BandWidth} with $X = 4 \, \%$. 

These findings further support the use of test-scalar QNMs as a quantitatively reliable proxy for the beyond-GR deviations of gravitational QNMs in alternative gravity theories, at least within the range of couplings currently allowed by observations.

\section{Application to a phenomenological metric: scalar QNMs of the Johannsen spacetime}
\label{Sec:JohannsenMetric}

The analyses performed in the previous sections suggest that, in well-motivated extensions of the Kerr metric, test-scalar QNMs closely reproduce the deviations from Kerr predicted by full gravitational perturbations, at a level of accuracy compatible with current ringdown measurements. This suggests that scalar modes can serve as a reliable and practical proxy for gravitational QNMs whenever the latter are difficult, or even impossible, to compute. This is particularly relevant for the classes of spacetime we consider in this section. 

In many proposed tests of the Kerr hypothesis, BH geometries are introduced at a purely metric level, without reference to an underlying field theory or a consistent set of dynamical equations. In such parametrically deformed metrics, gravitational perturbations cannot in general be derived from first principles, since the field equations governing the dynamics are unknown. By contrast, the dynamics of test fields --~and in particular minimally coupled scalar perturbations~-- remain well defined and straightforward to compute, especially if the background admits separability of curved spacetime D'Alembertian operator.

As a representative example, we consider a widely studied parametrically-deformed Kerr BH, the Johannsen metric~\cite{Johannsen:2013szh}. Although it is not a vacuum solution of any known theory beyond GR, it provides a controlled framework to parametrize deviations from the Kerr geometry and to assess their observational impact, for instance in tests of gravity based on BH imaging~\cite{EventHorizonTelescope:2019ths,EventHorizonTelescope:2022xqj}. As such, it offers an ideal arena to explore the complementarity between different observational probes of strong gravity, and in particular to assess how constraints from ringdown measurements compare with --~and potentially strengthen~-- those derived from BH shadow observations.

\subsection{General features of the metric}

The Johannsen metric is constructed to be free of pathological features outside the event horizon and to admit a Carter-like constant~\cite{Carter:1968rr}, unlike most beyond-Kerr models. This latter property --~also shared by the Kerr-Newman family~-- is crucial for simplifying the computation of observables. The existence of an additional constant of motion ensures complete separability of the Hamilton–Jacobi equation, allowing the geodesic equations to be written in first-order form. 

In the original construction~\cite{Johannsen:2013szh}, the Kerr metric is deformed according to these guiding principles through the introduction of scalar functions $f(r)$, $g(\theta)$, $A_i(r)$ ($i = 1, \, 2, \, 5$), and $A_j(\theta)$ ($j = 3, \, 4, \, 6$). Restricting each function to depend solely on either $r$ or $\theta$ provides the simplest parametrization of deviations from Kerr that preserves separability. Imposing asymptotic flatness and consistency with the current post-Newtonian constraints further reduces the number of independent functions to the $A_i(r)$'s and $f(r)$~\cite{Johannsen:2013szh}. The final metric components, written in Boyer-Lindquist coordinates, take the form
\begin{align}
    g_{tt} &= -\frac{\tilde \Sigma \left[\Delta-a^2 A_2(r)^2 \sin^2 \theta \right]}{\left[(r^2 + a^2)A_1(r) - a^2 A_2(r) \sin^2 \theta \right]^2}\, ;\nonumber\\
     g_{t\phi} &= -\frac{a\left[(r^2+a^2)A_1(r) A_2(r) - \Delta \right]\tilde \Sigma \sin^2 \theta}{\left[(r^2+a^2)A_1(r) - a^2 A_2(r) \sin^2 \theta \right]^2}\, ;\nonumber \\
      g_{rr} &= \frac{\tilde \Sigma}{\Delta A_5(r)}\, ;\nonumber \\
      g_{\theta \theta} &= \tilde \Sigma\, ;\nonumber \\
      g_{\phi \phi} &= \frac{\tilde \Sigma \sin^2 \theta \left[(r^2+a^2)^2 A_1(r)^2 -a^2 \Delta \sin^2 \theta \right]}{\left[(r^2+a^2)A_1(r) - a^2 A_2(r) \sin^2 \theta \right]^2} \, ,
      \label{eq:JohannsenMetric}
\end{align}
where 
\begin{align}
    \Delta & = r^2 -2Mr+a^2\, ; \nonumber \\
    A_1(r) & = 1 + \sum_{n=3}^\infty \alpha_{1n}\left(\frac{M}{r} \right)^n \, ; \nonumber \\
    A_2(r) & = 1+\sum_{n=2}^{\infty} \alpha_{2n}\left(\frac{M}{r} \right)^n\, ; \nonumber \\
    A_5(r) & = 1+\sum_{n=2}^{\infty} \alpha_{5n} \left(\frac{M}{r} \right)^n\, ; \nonumber \\
    \tilde \Sigma(r, \, \theta) & = r^2 + a^2 \cos^2 \theta + f(r)\, ; \nonumber \\
    f(r) & = \sum_{n=3}^\infty \epsilon_n \frac{M^n}{r^{n-2}}\, ,
    \label{eq:ModificationsKerrJohannsen}
\end{align}
with the $\alpha_{in}$'s and $\epsilon_n$'s constant deformation parameters. When all the latter vanish, the above metric reduces to the Kerr spacetime. 

Despite the presence of deformations, the locations of the horizons coincide with those of Kerr~\cite{Johannsen:2013szh}, since they are determined by $\Delta(r_\pm) = 0$, yielding $r_\pm = M \pm \sqrt{M^2-a^2}$. The deformation functions, thus, modify the exterior geometry while preserving the horizon structure. 

In what follows, we restrict our analysis to the lowest-order terms in the expansion~\eqref{eq:ModificationsKerrJohannsen}, namely
\begin{equation}
\begin{split}
    A_1(r) &= 1+\alpha_{13}\frac{M^3}{r^3}\, ; \qquad A_2(r) = 1 + \alpha_{22} \frac{M^2}{r^2}\, ;\\
    A_5(r) &= 1+\alpha_{52}\frac{M^2}{r^2}\, ; \hspace{0.9 cm} f(r)  = \epsilon_3 \frac{M^3}{r}\, .
    \label{eq:TruncatedModifyingFuncts}
\end{split}
\end{equation}
This truncation is motivated not only by computational simplicity, but also by phenomenological considerations: lower-order coefficients are expected to have the strongest impact on observables and are, therefore, the most amenable to observational constraints. Higher-order contributions could, in principle, be constrained sequentially, order by order, as observational precision improves. With the truncation~\eqref{eq:TruncatedModifyingFuncts}, the spacetime is characterized by six free parameters: $M$, $a$, $\alpha_{13}$, $\alpha_{22}$, $\alpha_{52}$, and $\epsilon_3$. However, when studying scalar test perturbations, we will impose an additional restriction on the modifying functions~\eqref{eq:ModificationsKerrJohannsen} to ensure separability of the Klein-Gordon equation. Under this condition, $f(r)$ is no longer independent, but can be expressed in terms of $A_1(r)$ and $A_2(r)$ [see \cref{eq:ConstraintfrSeparability} below]. Consequently, in the remainder of this work, we do not treat $\epsilon_3$ as an independent parameter and focus instead on constraints on $\alpha_{13}$, $\alpha_{22}$ and $\alpha_{52}$.

\subsubsection{Theoretical constraints on the parameters}
\label{subsubsec:TheoreticalConstraints}

We first discuss theoretical constraints on the deformation parameters, dictated by requiring regularity of the spacetime outside the horizon. In particular, we demand absence of both signature changes in the metric and closed time-like curves. 

To prevent signature changes, the metric components $g_{rr}$ and $g_{\theta \theta}$ must remain positive for $r > r_+$. This translates into the conditions $\tilde \Sigma >0$ and $A_5(r) >0$. Within the truncation~\eqref{eq:TruncatedModifyingFuncts}, the last requirement implies $\alpha_{52}>0$. Furthermore, preserving the Lorentzian signature requires the metric determinant to remain negative. This condition yields a lower bound on $\alpha_{22}$~\cite{Johannsen:2013szh}
\begin{equation}
    \alpha_{22} >-\frac{r^2_+}{M^2}\, .
    \label{eq:Boundalpha22}
\end{equation}
Finally, requiring absence of closed time-like curves outside the outer horizon, i.e., $g_{\phi\phi} > 0$, yields the lower bound on $\alpha_{13}$~\cite{Johannsen:2013szh}: 
\begin{equation}
    \alpha_{13}>-\frac{r_+^3}{M^3}\, .
    \label{eq:Boundalpha13}
\end{equation}
Consistent with these bounds, in this work we restrict our analysis to $\alpha_{52} > 0, \, \alpha_{13} > -1, \, \alpha_{22} > -1$. The latter lower bounds correspond to maximally-rotating BHs with $a = M$.

\subsubsection{Observational constraints}
\label{sec:ObservationalConstraints}

We then turn to reviewing some observational constraints on the deformation parameters. 

One possible avenue to constrain the metric \eqref{eq:JohannsenMetric} is by comparing its predicted BH shadow with EHT observations~\cite{Johannsen:2010ru,Broderick:2013rlq,Medeiros:2019cde,EventHorizonTelescope:2022xqj} \footnote{Interpretations of beyond-Kerr shadows must be handled with care, since geometries with strong non-Kerr multipolar structure can nevertheless produce shadows nearly indistinguishable from Kerr~\cite{Glampedakis:2023eek}.}. As shown in~\cite{Johannsen:2013vgc}, the properties of the light ring depend only on $\alpha_{13}$ and $\alpha_{22}$, and are independent of $\alpha_{52}$. This occurs because $A_5(r)$ modifies only the $g_{rr}$ metric component and therefore does not affect equatorial geodesic motion at the light ring (see also \cref{subsubsec:EikonalApproxJohannsen}).

A detailed analysis of shadow morphologies in the metric~\eqref{eq:JohannsenMetric} was performed in Ref.~\cite{Medeiros:2019cde}. It was found that $\alpha_{22}$ primarily controls deviations from circularity, with larger values increasing distortions. By contrast, $\alpha_{13}$ mainly shifts the light-ring radius and hence the overall angular scale of the shadow; increasing $\alpha_{13}$ tends to enlarge the shadow while simultaneously making it more circular. 

In practice, however, extracting tight bounds on $\alpha_{13}$ and $\alpha_{22}$ is hindered by several factors: the structural complexity of the metric, astrophysical uncertainties in the emission model, and instrumental limitations. Current EHT constraints are driven predominantly by the angular diameter of the emission ring, which is the most robust and model-independent observable extracted from the data. This quantity depends only weakly on the black hole spin and inclination~\cite{Johannsen:2010ru,Johannsen:2013vgc,Medeiros:2019cde}. Moreover, the EHT analyses of both M87~\cite{EventHorizonTelescope:2019ths} and SgrA$^\ast$~\cite{EventHorizonTelescope:2022xqj} are designed to remain as astrophysically agnostic as possible, minimizing assumptions about plasma dynamics, emission mechanisms, and radiative transfer.

This strategy strengthens constraints on parameters that primarily affect the global geometric scale, such as $\alpha_{13}$,  while weakening bounds on parameters that mainly modify the shadow shape, such as $\alpha_{22}$. Distortions induced by $\alpha_{22}$ are highly degenerate with astrophysical effects, including disk thickness, system inclination, and details of the emission profile. These effects can mimic or obscure small geometry-induced deviations from circularity. In addition, current limitations in angular resolution and $(u,v)$ coverage may introduce asymmetries or artifacts, making it challenging to isolate percent-level deviations in shape independently of emission modeling (see, e.g., Ref.~\cite{EventHorizonTelescope:2025whi} and references therein).

As a result, $\alpha_{13}$ is relatively tightly constrained: $-3.6 < \alpha_{13} < 5.9$ for M87~\cite{EventHorizonTelescope:2020qrl}, and $-3.1 \lesssim \alpha_{13} \lesssim 1.5$ for SgrA$^\ast$~\cite{EventHorizonTelescope:2022xqj}, whereas shadow-based bounds on $\alpha_{22}$ remain comparatively weak~\cite{EventHorizonTelescope:2019ths,Tiede:2022bdd,Tiede_2024}.

In this context, comparisons with other strong-field probes, such as QNMs measured in GW ringdowns, are particularly valuable. These observables provide constraints on the deformation parameters that are independent of, and complementary to, those derived from imaging data.

Since the Johannsen metric is not a vacuum solution of any known gravitational theory, gravitational QNMs cannot be computed from first principles~\footnote{One possibility is to interpret the metric as a GR solution in the presence of exotic matter. However, this requires finding a covariant stress-energy tensor that correctly reduces to the one sourcing the solution. Since this is challenging, especially for spinning geometries, previous approaches have neglected matter perturbations in the linearized gravitational equations~\cite{Guo:2024mmq,Wu:2025obg}.}.
Test-field perturbations, instead, can be consistently studied on the fixed background. In light of the results established in the previous sections, test-scalar QNMs offer a reliable quantitative proxy for the deviations of the gravitational spectrum away from the Kerr case, within current observational accuracy. We therefore employ them in what follows to assess ringdown phenomenology in this parametrically deformed Kerr spacetime.

\subsection{Massless Klein-Gordon Equation}

Massless scalar test perturbations are governed by the Klein-Gordon equation
\begin{equation}
    \Box \Psi = \frac{1}{\sqrt{-g}}\partial_\mu \left(\sqrt{-g}\, g^{\mu\nu} \, \partial_\nu \Psi \right) = 0\, ,
    \label{eq:KGequation}
\end{equation}
where $\Psi = \Psi(t, \, r, \, \theta, \, \phi)$. A standard strategy consists in working in the frequency domain, assuming harmonic dependence on $t$ and $\phi$ (ensured by the stationarity and axisymmetry of the background, respectively), together with separability in $r$ and $\theta$. Specifically, we adopt the ansatz
\begin{equation}
    \Psi(t, \, r, \, \theta, \, \phi) = \ee^{-\ii \omega t} \ee^{\ii m \phi} \, R(r) \, S(\theta)\, ,
    \label{eq:AnsatzPsi}
\end{equation}
where $\omega$ denotes the mode frequency, while $m$ is, again, the azimuthal number. In the Kerr spacetime, the Klein-Gordon equation is completely separable into independent radial and angular equations, yielding the scalar Teukolsky system. Such separability is not generically preserved in beyond-Kerr geometries, even when a Carter-like constant exists. In general, deviations from Kerr introduce couplings between radial and angular variables, thereby obstructing separability.

For the Johannsen metric~\eqref{eq:JohannsenMetric}, separability can nevertheless be achieved by restricting to a specific subclass of models satisfying the condition~\cite{Konoplya:2018arm} 
\begin{equation}
    f(r) = (r^2 + a^2)\left[\frac{A_1(r)}{A_2(r)}-1 \right]\, .
    \label{eq:ConstraintfrSeparability}
\end{equation}
In Appendix~\ref{App:KGJohannsen}, we provide a detailed derivation of the resulting radial and angular equations, explicitly showing how the constraint~\eqref{eq:ConstraintfrSeparability} guarantees separability. The final equations take the form
\begin{equation}
\begin{split}
    \frac{1}{\sin\theta} \partial_\theta \left[\sin \theta \partial_\theta S(\theta) \right] &+ \biggl(a^2 \omega^2 \cos^2 \theta \\
    &-\frac{m^2}{\sin^2 \theta} + \Lambda_{\ell m} \biggr) S(\theta) = 0\, ;
    \label{eq:Teu1}
\end{split}
\end{equation}
\begin{equation}
A_2 \sqrt{A_5} \partial_r \left(\frac{\sqrt{A_5}}{A_2} \Delta \, R' \right) + \mathcal{V}(r) R(r) = 0\, ,
\label{eq:Teu2}
\end{equation}
where we have defined
\begin{equation}
\begin{split}
    \mathcal{V}(r) \equiv & \frac{(r^2+a^2)^2}{\Delta} \omega^2 A_1^2-\frac{2am \omega}{\Delta}\left[(r^2+a^2) A_1 A_2 - \Delta \right]\\
    &+ \frac{a^2m^2}{\Delta}A_2^2-\left(\Lambda_{\ell m} + a^2 \omega^2\,  \right)\, ,
\end{split}
\label{eq:EffectiveVKG}
\end{equation}
while $\Lambda_{\ell m}$ denotes the separation constant.
In the limit where all deformation functions $A_i(r)$ reduce to unity, the standard scalar Teukolsky equations in Kerr are recovered.

The system remains coupled through the shared eigenvalues $\Lambda_{\ell m}$ and frequencies $\omega$, which enter both equations. In the following, we outline the strategy used to determine these quantities, focusing in particular on the QNM spectrum.

\subsubsection{Boundary conditions}

We begin by specifying the boundary conditions. 

The angular equation coincides with its Kerr counterpart, since the deformation functions depend only on $r$ and do not affect the angular sector. Consequently, the boundary conditions are identical to those in the Kerr case~\cite{Leaver:1985ax}: regularity of $S(\theta)$ at the regular singular points $\theta = 0$ and $\theta = \pi$. 

The radial equation, by contrast, is modified and requires a dedicated analysis. Since we are interested in the ringdown of the spacetime~\eqref{eq:JohannsenMetric}, we interpret $\omega$ as the QNM frequency. The corresponding boundary conditions select solutions that are purely outgoing at spatial infinity and purely ingoing at the outer horizon $r_+$. At large radial distances, the deformation functions~\eqref{eq:TruncatedModifyingFuncts} introduce corrections that decay as $\mathcal{O}(r^{-\gamma})$ with $\gamma \geq 2$. Consequently, the asymptotic structure of the radial equation coincides with that of Kerr, and the purely outgoing solution at infinity behaves as
\begin{equation}
    R(r) \sim \ee^{\ii \omega r} \, r^{-1+2\ii M \omega } \qquad \text{for} \quad r \to \infty\, .
    \label{eq:BehaviorInfinity}
\end{equation}
Near the outer horizon $r = r_+$, deviations from the Kerr metric become relevant. Solving the radial equation in the vicinity of $r_+$, we obtain the following behavior for the purely ingoing mode:
\begin{equation}
    R(r) \sim (r-r_+)^{-\ii \sigma} \qquad \text{for} \quad r \sim r_+\, ,
\end{equation}
with $\sigma$ given by
\begin{align}
    \sigma \equiv \frac{-a m r_+ \left(r_+^2 + \alpha_{22} M^2\right)+\omega  \left(r_+^2+ a^2\right) \left(\alpha_{13} M^3+r_+^3\right)}{r_+^2 (r_+ -r_-) \sqrt{r_+^2 + \alpha_{52} M^2}}\, .
    \label{eq:sigmaboundarydef}
\end{align}
In the Kerr limit, i.e., $\alpha_{13}, \, \alpha_{22}, \, \alpha_{52} \to 0$, the above expression reduces to the well-known result~\cite{Leaver:1985ax}
\begin{equation}
    \sigma = \frac{2M \omega r_+ -a m}{2 \sqrt{M^2-a^2}} \, .
\end{equation}

\subsubsection{Solution strategies for the modified Teukolsky equations}
\label{subsubsec:SolutionStrategy}

In this subsection, we describe the semi-analytical methods employed to solve the modified Teukolsky equations and to determine simultaneously the separation constants $\Lambda_{\ell m}$ and the QNM frequencies $\omega$.

The procedure begins with an ansatz for the angular function $S(\theta)$ that explicitly incorporates the correct behavior at the two regular singular points $\theta = 0$ and $\theta = \pi$, together with a Frobenius series expansion about one of the poles. Substituting the ansatz into \cref{eq:Teu1} yields a three-term recurrence relation for the series coefficients. Requiring convergence over the full angular domain leads to a continued-fraction condition on the recurrence relation (see~\cite{Leaver:1985ax} for further details). Solving this condition numerically provides the separation constants $\Lambda_{\ell m}$ for given $\omega$.

The radial equation, although modified with respect to the Kerr case, retains the same singularity structure: two regular singular points at $r=r_\pm$ and an irregular singular point at infinity. We therefore factor out the asymptotic behavior and perform a Frobenius expansion about $r = r_+$. The ansatz takes the form
\begin{align}
    R(r) &= \ee^{\ii \omega r} (r-r_-)^{\lambda} (r-r_+)^\kappa Z(r)\, ;\nonumber\\
    Z(r) & \equiv \sum_{n= 0}^{\infty} d_n \left(\frac{r-r_+}{r-r_-} \right)^n\, ,
    \label{eq:ansatzRadial}
\end{align}
where the constants $\kappa$ and $\lambda$ are fixed by the boundary conditions. Regularity and ingoing behavior at $r_+$ implies $\kappa \equiv -\ii \sigma$, with $\sigma$ defined in \cref{eq:sigmaboundarydef}. Imposing the asymptotic behavior at infinity~\eqref{eq:BehaviorInfinity}, i.e., $R(r) \sim r^{-1+2\ii M\omega}$, yields $\lambda = -1 + 2\ii M\omega - \kappa$. In the following, we adopt Leaver’s conventions~\cite{Leaver:1985ax} and set the BH mass to $M = 1/2$. Introducing the compactified radial coordinate $x \equiv (r-r_+)/(r-r_-)$ and substituting the ansatz~\eqref{eq:ansatzRadial} into \cref{eq:Teu2} leads to an infinite linear system for the coefficients $d_n$, governed by a $13$-term recurrence relation
\begin{equation}
\begin{split}
   &c_0(0) d_1 + c_1(0) d_0 = 0\, ;\\
   &c_0(1) d_2 + c_1(1)d_1 + c_2(1) d_0 = 0\, ; \\
   &c_0(2) d_3 + c_1(2) d_2 + c_2(2) d_1 + c_3(2) d_0 = 0\, ;\\
   &c_0(3) d_4 + c_1(3) d_3 + c_2(3) d_2 + c_3(3) d_1 + c_4(3) d_0 = 0\, ;\\
   &\dots\\
   &c_0(n) d_{n+1} + c_1(n) d_n + c_2(n) d_{n-1} + c_3(n) d_{n-2} \\
   &+ c_4(n) d_{n-3} + c_5(n) d_{n-4} + c_6(n) d_{n-5} \\
    &+ c_7(n) d_{n-6} + c_8(n) d_{n-7}+ c_9(n) d_{n-8} \\
    & + c_{10}(n) d_{n-9} + c_{11}(n) d_{n-10} + c_{12}(n) d_{n-11} = 0\, ,\\
    & (n = 11, 12, \dots)
\end{split}
\label{eq:RecurrenceJohannsen}
\end{equation}
whose coefficients depend on $a$, $\alpha_{13}$, $\alpha_{22}$, $\alpha_{52}$, $n$, $\omega$ and $\Lambda_{\ell m}$. Their explicit expressions are lengthy and will not be reported here. They are, instead, provided in an ancillary \textsc{Mathematica} notebook available at the link~\cite{Bitbucket_repo}.

The system can be recast as a matrix equation admitting nontrivial solutions only when the associated determinant --~known as the Hill determinant~-- vanishes~\cite{Majumdar:1989tzg,Leaver:1990zz}. In practice, the infinitely-dimensional determinant is approximated by truncating the matrix to a finite size $N\times N$. The zeros of the truncated determinant provide numerical approximations to the QNM frequencies, and convergence is assessed by increasing $N$ until the values of the obtained roots stabilize. 

To couple the radial and angular sectors, we implement an iterative procedure. Starting from a trial value of $\Lambda_{\ell m}$, the truncated Hill determinant is solved numerically for $\omega$ using a root-finding algorithm. The resulting frequency is then inserted into the angular equation, which is solved via Leaver’s continued-fraction method to update $\Lambda_{\ell m}$. This process is iterated until both $\omega$ and $\Lambda_{\ell m}$ converge to stable values.

As a validation test, we set all deformation parameters to zero and verify that the scalar QNMs of Schwarzschild and Kerr are reproduced with high accuracy. For the radial equation, we truncate the determinant at $N=200$ and compute frequencies with $30$-digit precision. The angular continued fraction is truncated after 130 terms, and the iteration is performed up to seven times (although fewer iterations are typically sufficient). With this setup, Schwarzschild and Kerr QNM frequencies are reproduced up to the $14$th significant digit.

\subsection{QNMs of the Johannsen metric and complementarity with shadow constraints}
\label{subsec:Results}

In this section, we present the results of scalar test QNMs in the Johannsen spacetime and analyze the impact of the individual deformation parameters on the mode frequencies. 
To isolate the role of each modification, we vary one deviation parameter at a time while setting the others to zero. This strategy allows us to disentangle their respective effects and to perform a direct comparison with the Kerr case. In Appendix~\ref{app:MixedConstantsJohannses} we complement this analysis with an illustrative multi-parameter study, where the impact of simultaneously varying two deformation parameters, $\alpha_{13}$ and $\alpha_{22}$, is investigated. The results are displayed as functions of the dimensionless spin $a/M$ and contrasted with the scalar QNMs of Kerr. 

We emphasize that, in general, multiple deviation parameters may be simultaneously non-zero and correlated, and deriving robust observational constraints on the underlying spacetime geometry would therefore require a joint Bayesian inference in which all relevant parameters are varied simultaneously. At the same time, it has been pointed out that valuable tests of GR can also be constructed by investigating whether only specific subsets of deviation parameters are allowed to differ from their GR values, rather than assuming from the outset that all of them are non-zero~\cite{Li:2011cg,Li:2011vx}. Although our analysis is not statistical in nature, it follows a similar philosophy by isolating the physical effect of each metric deformation. Additionally, the results of Appendix~\ref{app:MixedConstantsJohannses} show that allowing more than one deformation parameter to vary simultaneously does not qualitatively change the main conclusions obtained from the single-parameter analysis.

While the abovementioned analyses represent a natural extension of our work, our objective here is to establish a direct correspondence between each individual metric deformation, its impact on the geometry of the underlying photon ring, and the resulting shift in the QNM spectrum, rather than to derive observational constraints on the Johannsen spacetime. More generally, a finite set of measured complex QNM frequencies cannot uniquely reconstruct an arbitrary spacetime metric or perturbation potential. Meaningful inference therefore requires a restricted theory-specific or metric-specific parametrization, possibly supplemented by independent observables such as BH shadow measurements.

To assess the potential observational significance of the deviations, we also include reference bands around the Kerr frequencies. As justified in the previous sections, within current measurement accuracy we can trade the gravitational modes with their scalar counterpart. Therefore, the width of the reference bands is chosen to reflect current constraints from gravitational QNM measurements. As a benchmark, we refer again to GW250114~\cite{LIGOScientific:2025rid}. As explained in \cref{Sec:TradingQNMdeviations}, for this event, deviations of the $(0,2,2)$ gravitational mode from the Kerr prediction were constrained at the level of approximately $4\, \%$ for the real part and $10\%$ for the imaginary part of the frequency~\cite{LIGOScientific:2025wao}. Thus, even in the scalar case, these bands provide a useful reference for assessing which regions of parameter space could be probed/excluded by present-day observations.

\subsubsection{$\alpha_{22} = \alpha_{52} = 0$}
\label{subsubsec:A1}

We first consider the simplest scenario in which $A_2$ and $A_5$ are set to unity, so that $\alpha_{13}$ is the only nonvanishing deformation parameter. In this case, \cref{eq:Teu2} simplifies considerably: the differential operator (the first term on the left-hand side) retains exactly the same structure as in Kerr, and the modifications enter only the effective potential~\eqref{eq:EffectiveVKG}.

\begin{figure}
    \centering
    \includegraphics[width=\linewidth]{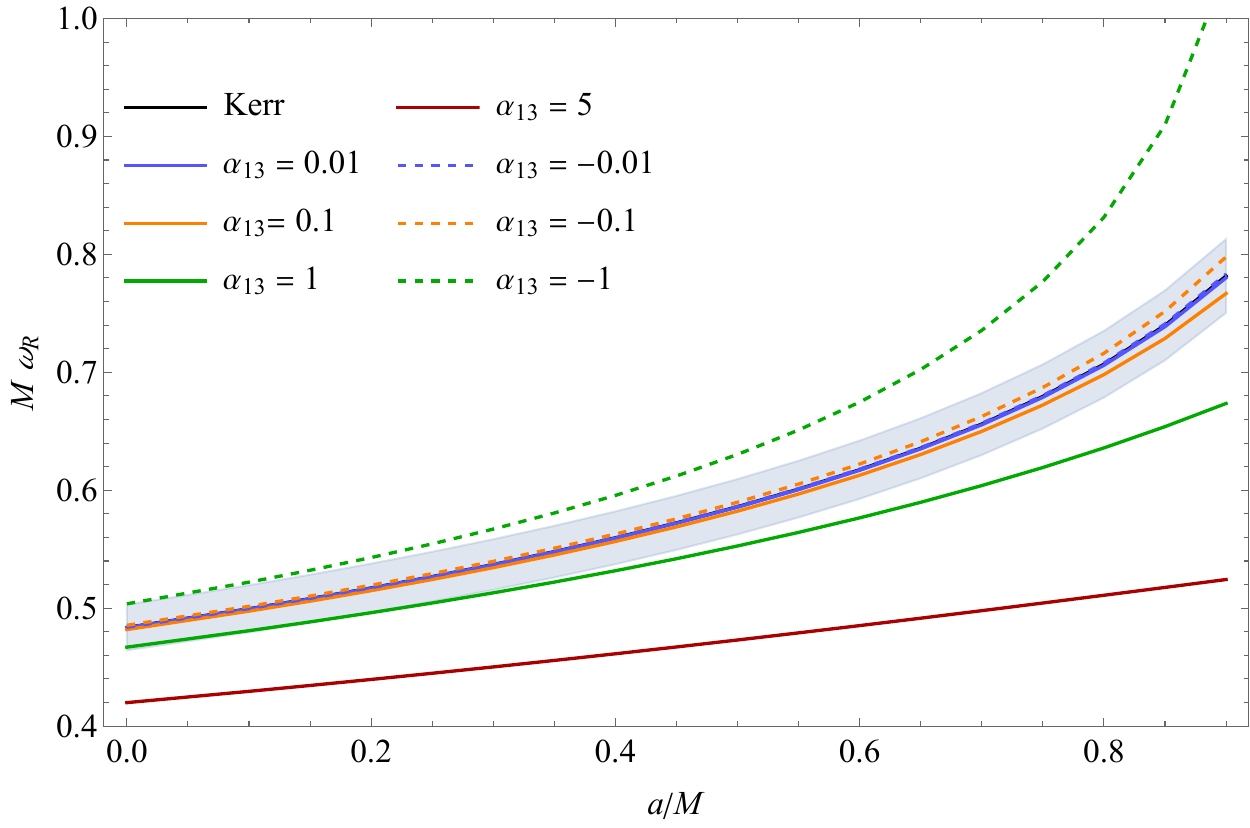}\hfill
    \includegraphics[width=\linewidth]{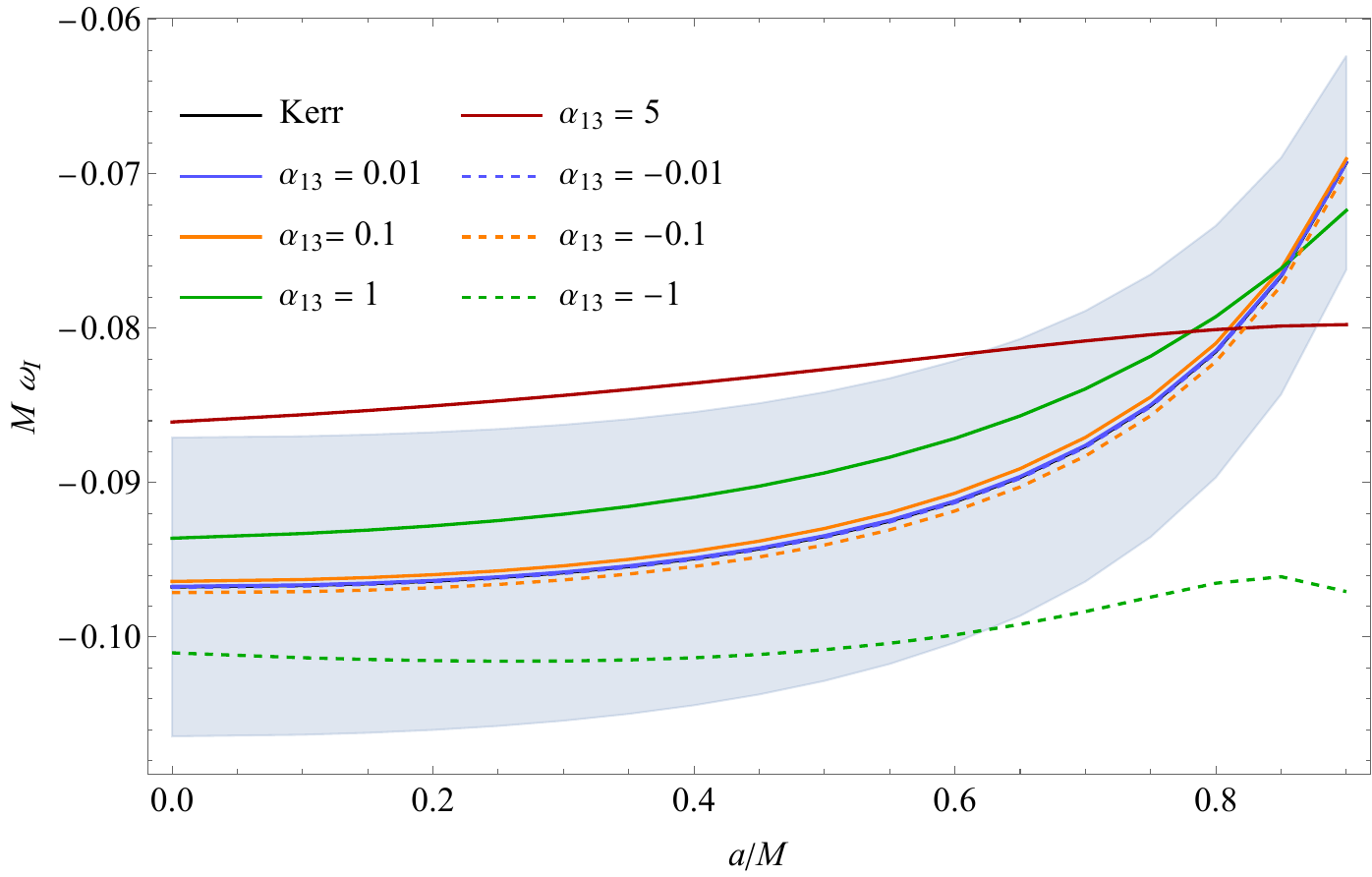}
        \caption{Real (top) and imaginary (bottom) parts of the QNM frequencies for the Johannsen metric \eqref{eq:JohannsenMetric} with $A_2 = A_5 = 1$. Results are shown as functions of the dimensionless spin $a/M$ for several values of the deformation parameter $\alpha_{13}$. Solid and dashed curves correspond to positive and negative values of $\alpha_{13}$, respectively. The Kerr scalar QNMs (black solid line) are also shown for comparison, but are nearly indistinguishable from the curves with $|\alpha_{13}|\sim 10^{-2}$. The blue shaded regions denote the $\pm 4\%$ and $\pm 10\%$ bands around the Kerr results for the real and imaginary parts, respectively.}
    \label{fig:JohannsenFirstCase}
\end{figure}

The results of the real and imaginary parts of the QNM frequencies are reported in \cref{fig:JohannsenFirstCase}.  
We notice that increasing positive values of $\alpha_{13}$ produce smaller real and larger imaginary parts. Conversely, increasingly negative values of $\alpha_{13}$ lead to larger real parts and to imaginary parts with larger absolute values. 

For $|\alpha_{13}| \gtrsim 1$, the real part of the frequency deviates by more than $4\, \%$ from the Kerr prediction over a significant portion of the spin range. Such configurations would therefore be potentially disfavored by current ringdown observations. These limits are qualitatively consistent with the shadow-based constraints discussed in \cref{sec:ObservationalConstraints}, and suggest that ringdown measurements could provide competitive --~possibly even stronger~--bounds on $\alpha_{13}$. 

\subsubsection{$\alpha_{13} = \alpha_{52} = 0$}
\label{subsubsec:A2}

\begin{figure}
    \centering
    \includegraphics[width=\linewidth]{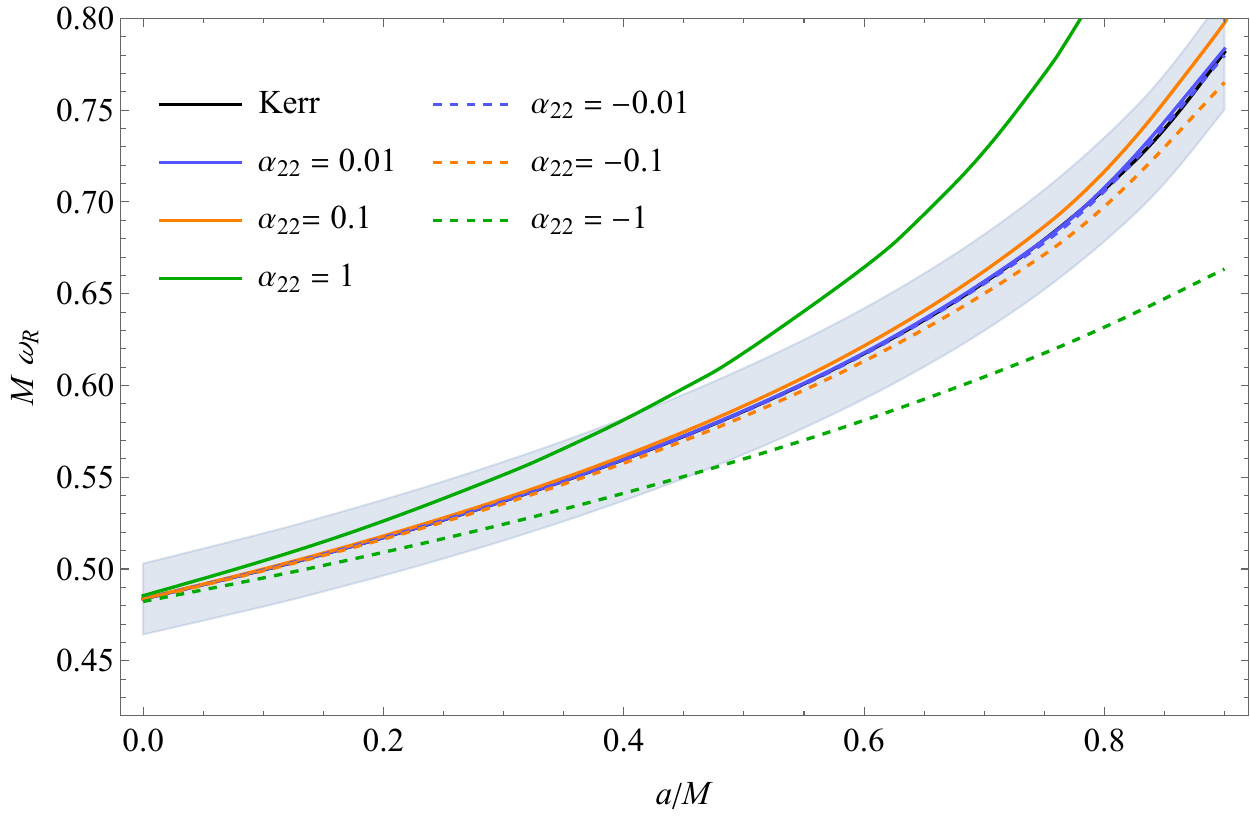}\hfill
    \includegraphics[width=\linewidth]{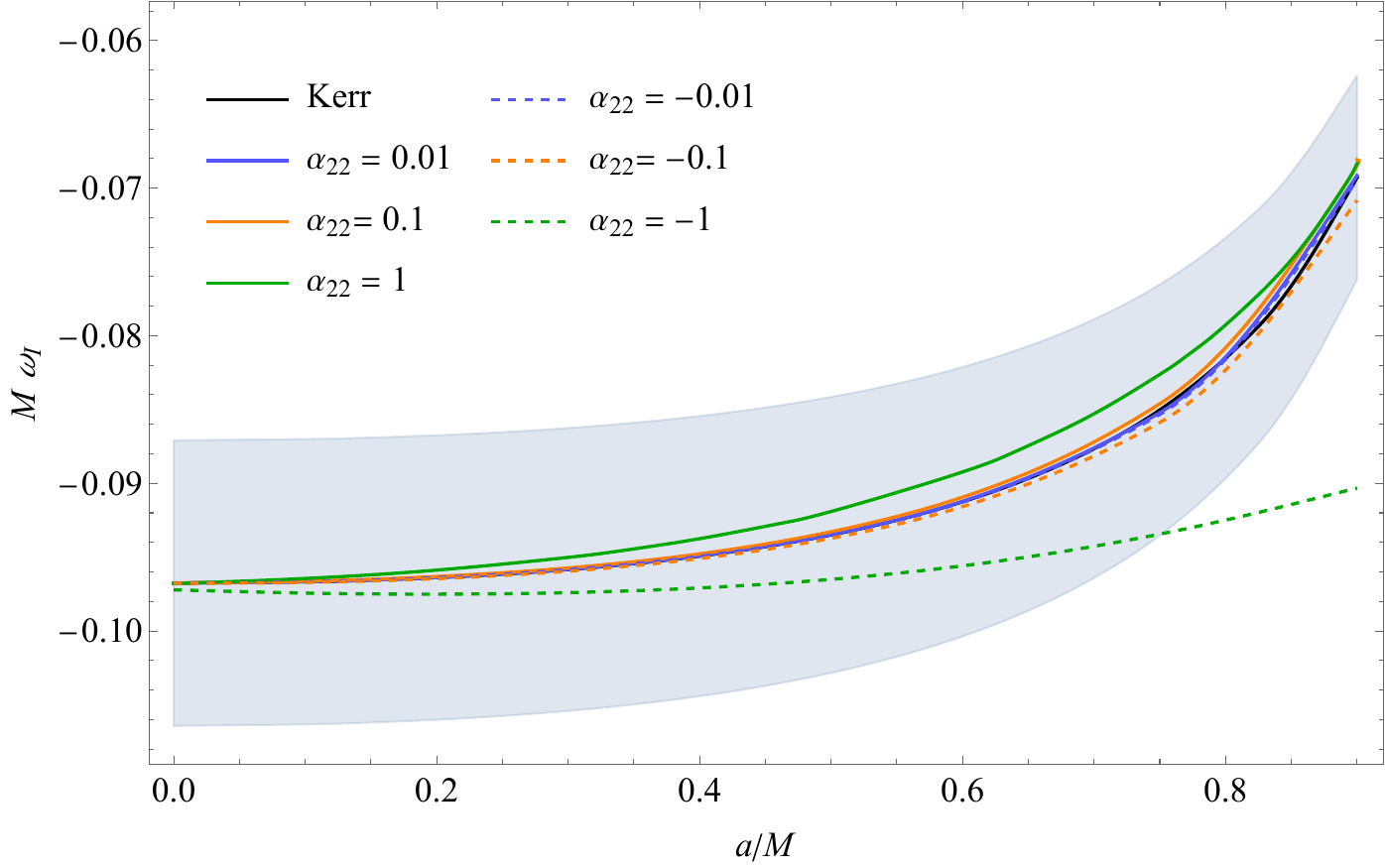}
        \caption{Real (top) and imaginary (bottom) parts of the QNM frequencies for the Johannsen metric \eqref{eq:JohannsenMetric} with $A_1 = A_5 = 1$. Results are shown as functions of the dimensionless spin $a/M$ for several values of the deformation parameter $\alpha_{22}$. Solid and dashed curves correspond to positive and negative values of $\alpha_{22}$, respectively. The Kerr scalar QNMs (black solid line) are also displayed, but are nearly indistinguishable from the curves with $|\alpha_{22}|\sim 10^{-2}$. The blue shaded regions denote the $\pm 4\%$ and $\pm 10\%$ bands around the Kerr results for the real and imaginary parts, respectively.}

    \label{fig:JohannsenA2}
\end{figure}

We now consider the case in which only $\alpha_{22}$ is nonzero. The corresponding QNM frequencies are shown in \cref{fig:JohannsenA2}.
The imaginary parts exhibit a trend similar to that found in the $\alpha_{13} \neq 0$ case: increasing positive values of $\alpha_{22}$ lead to larger imaginary parts, while negative values lead to a decrease. The real parts, however, show the opposite trend compared to the previous subsection. Increasing positive values of $\alpha_{22}$ produce larger real frequencies, whereas negative values result in decreasing imaginary parts.

We further find that configurations with $|\alpha_{22}| \gtrsim 1$ yield real frequencies that exceed the $4\%$ band around the Kerr values for spins $a/M \gtrsim 0.4$, suggesting that such deviations could also be excluded by GW observations, at variance with BH shadow tests
  
\subsubsection{$\alpha_{13} = \alpha_{22} = 0$}
\label{subsubsec:A5}

\begin{figure}
    \centering
    \includegraphics[width=\linewidth]{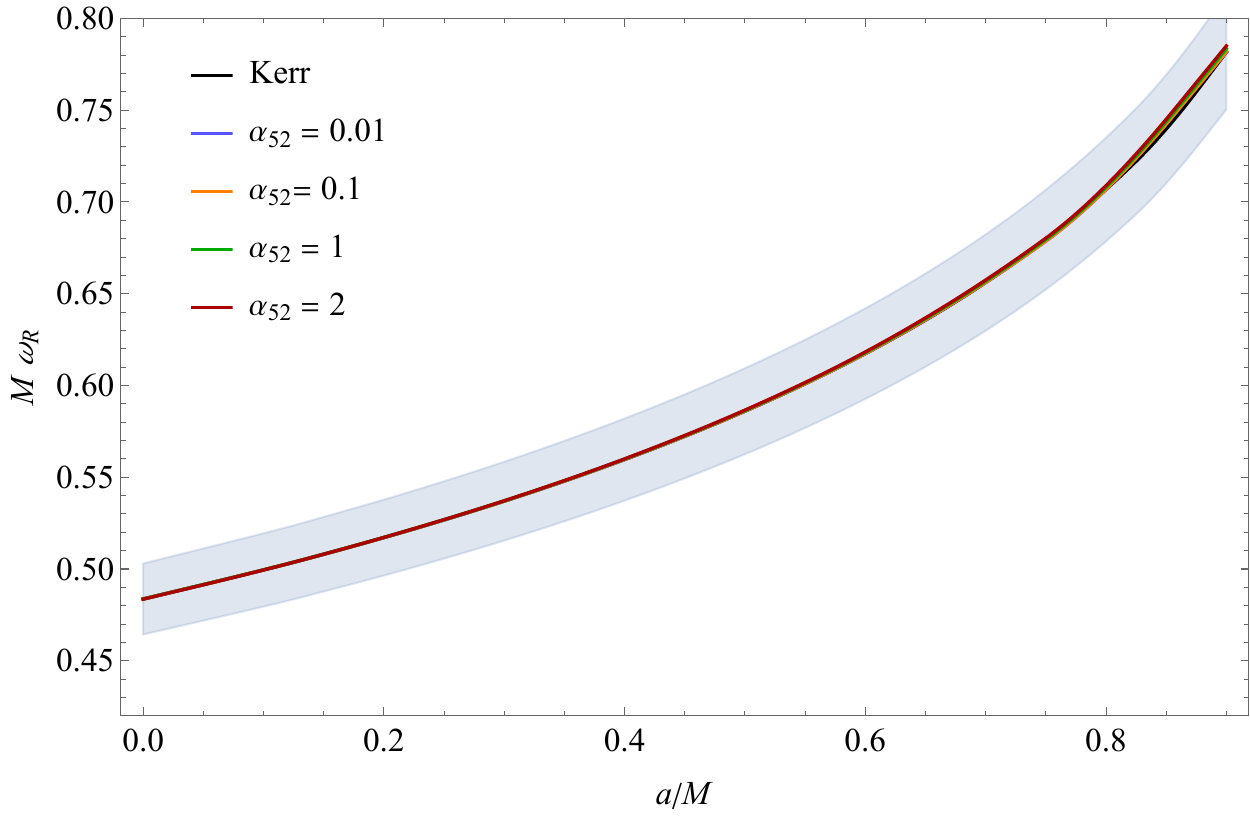}\hfill
    \includegraphics[width=\linewidth]{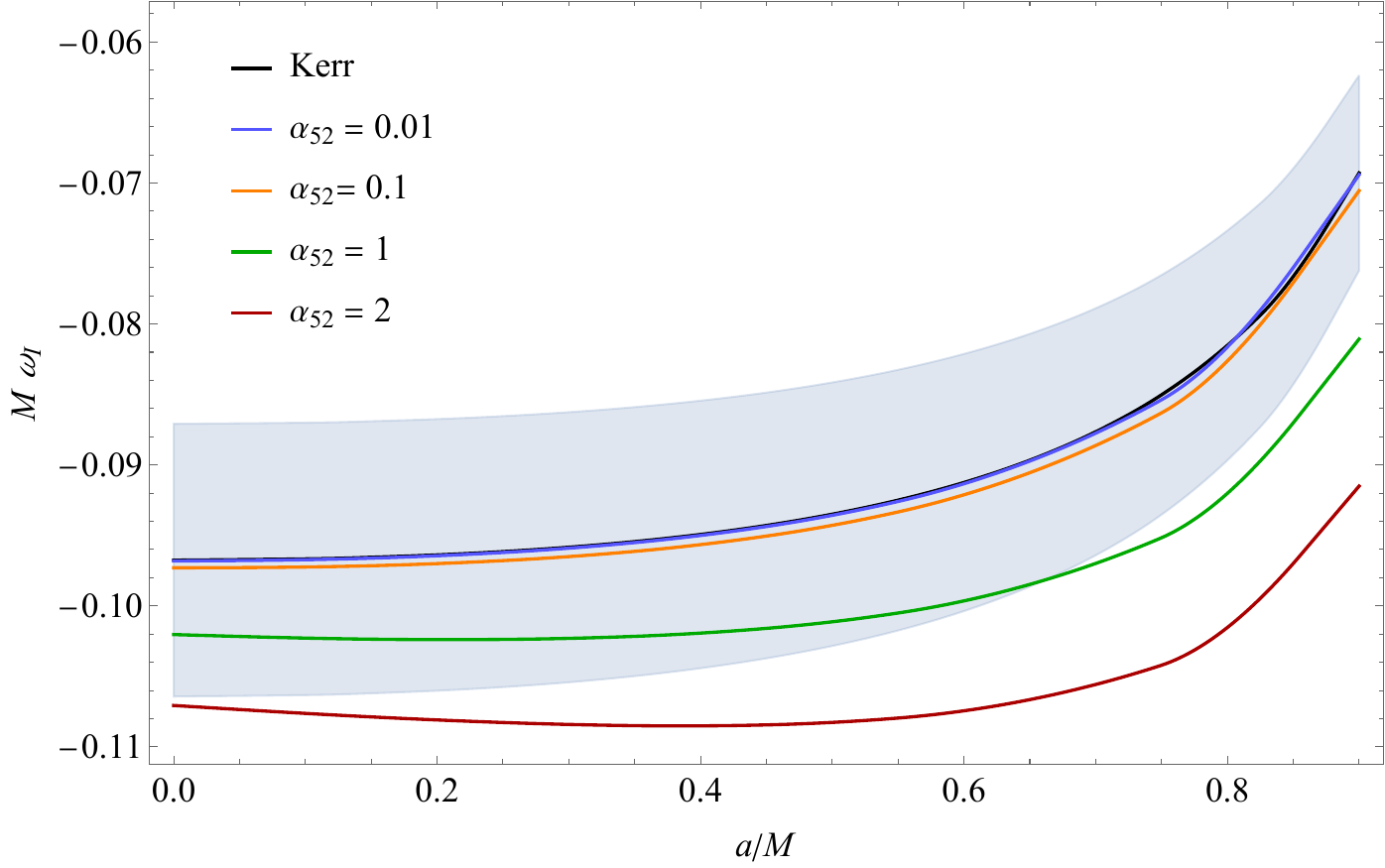}
        \caption{Real (top) and imaginary (bottom) parts of the QNM frequencies for the Johannsen metric \eqref{eq:JohannsenMetric} with $A_1 = A_2 = 1$. The frequencies are shown as functions of the dimensionless spin $a/M$ for several values of the deformation parameter $\alpha_{52}$. The scalar QNMs of the Kerr spacetime (black solid line) are included for comparison. The real parts of the modified frequencies are nearly indistinguishable from the Kerr ones. For the imaginary parts, the Kerr curve and the case with $\alpha_{52}\sim 10^{-2}$ are almost overlapping. The blue shaded regions denote the $\pm 4\%$ and $\pm 10\%$ bands around the Kerr results for the real and imaginary parts, respectively.}
    \label{fig:JohannsenA5}
\end{figure}

Finally, we examine the effect of the deformation parameter $\alpha_{52}$. The results are shown in \cref{fig:JohannsenA5}.
In contrast to the previous cases, the real part of the frequency exhibits only a very mild dependence on $\alpha_{52}$ and remains nearly indistinguishable from the Kerr prediction across the explored parameter range. The imaginary part, however, is significantly more sensitive. Increasing $\alpha_{52}$ systematically increases the magnitude of the absolute value. In particular, for $\alpha_{52} \gtrsim 2$, the imaginary parts depart from the Kerr value by more than $10\, \%$, placing such deviations outside the current observational band. This suggests that sufficiently large values of $\alpha_{52}$ could already be disfavored by ringdown measurements, even though this parameter does not affect the real part appreciably.

\subsubsection{Interpretation within eikonal approximation}
\label{subsubsec:EikonalApproxJohannsen}

We now turn to an interpretation of the above results within the eikonal approximation. This perspective is particularly valuable in the present context, given the central role of the light ring in EHT observations and the sensitivity of its properties to the underlying spacetime parameters (see \cref{sec:ObservationalConstraints}). As reviewed above, $\alpha_{13}$ and $\alpha_{22}$ affect the light-ring size and eccentricity, respectively, whereas $\alpha_{52}$ has no impact on its geometry. This distinction highlights the complementarity between shadow and ringdown measurements: whereas certain deviations from Kerr can be constrained through BH shadows, others may manifest primarily in the QNM damping rates and be accessible only via ringdown observations. Additionally, even though the eikonal approximation is not here used to explicitly compute the full QNM spectrum, it provides a valuable interpretative tool to understand the trends observed in the numerical results discussed in the previous sections. 

We first recall the result of Ref.~\cite{Johannsen:2013vgc}, according to which $\alpha_{52}$ has no impact on the location of the light ring. Considering equatorial photon motion at $\theta = \pi/2$ and using the metric components in~\eqref{eq:JohannsenMetric}, the effective potential~\eqref{eq:AppVEff} can be evaluated. The light-ring radius and the ratio $\mathcal{L}/\mathcal{E} \equiv \bar{\mathcal{L}}$ among the geodesic constants of motion are then obtained by solving the system~\eqref{eq:AppLR}, which reduces to (see also Ref.~\cite{Johannsen:2013vgc})
\begin{equation}
    \mathcal{F}(r_\text{LR}, \, \bar{\mathcal{L}}) = \frac{\dd}{\dd r} \mathcal{F}(r, \bar{\mathcal{L}})\biggr|_{r = r_\text{LR}} = 0\, ,
\end{equation}
where we have defined
\begin{equation}
    \mathcal{F}(r, \, \bar{\mathcal{L}}) \equiv \left[A_1 \left(a^2+r^2\right)-a\, A_2\, \bar{\mathcal{L}} \right]^2-(a-\bar{\mathcal{L}} )^2 \Delta \, . 
\end{equation}
Importantly, $\mathcal{F}$ does not depend on $A_5$, confirming that the light-ring position is independent of $\alpha_{52}$. Since the orbital frequency at the light ring depends only on the metric components that enter equatorial motion, it follows that the real part of the QNM frequencies in the eikonal regime is unaffected by $\alpha_{52}$. This is in perfect agreement with the numerical trends shown in the upper panel of \cref{fig:JohannsenA5}. By contrast, the second derivative of the effective potential --~entering the Lyapunov exponent and thus determining the imaginary part of the QNMs in the eikonal approximation [see \cref{eq:AppImaginaryQNM}]~-- is sensitive to $A_5$. This explains why $\alpha_{52}$ predominantly affects the imaginary frequencies, as seen in the lower panel of \cref{fig:JohannsenA5}.

\begin{figure}
    \centering
    \includegraphics[width=0.8\linewidth]{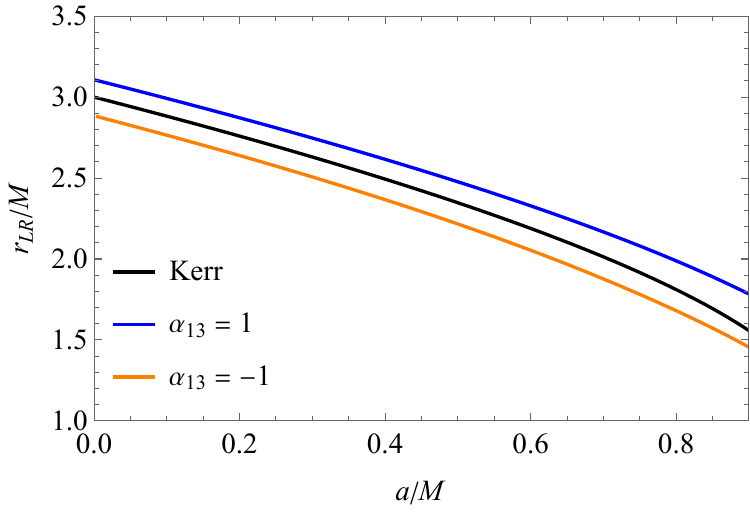}\hfill
    \includegraphics[width=0.8\linewidth]{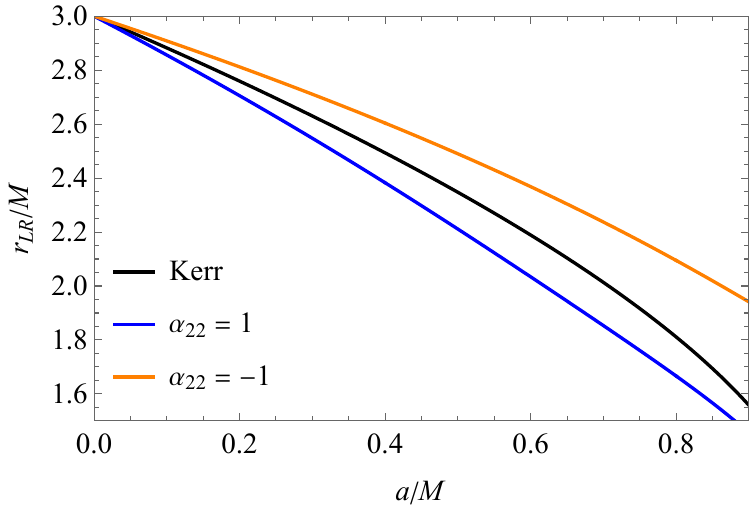}
    \caption{Top panel: light-ring radius as a function of $a/M$ for different values of $\alpha_{13}$, with $\alpha_{22} = 0$. The black line corresponds to $\alpha_{13} = 0$ (Kerr), while the blue and orange curves correspond to $\alpha_{13} = 1$ and $-1$, respectively. 
    Bottom panel: same as the top panel, but with $\alpha_{13} = 0$ and varying $\alpha_{22}$. The color code is the same.}
    \label{fig:JohannsenLR}
\end{figure}

The eikonal picture also clarifies the opposite trends observed when varying $\alpha_{13}$ or $\alpha_{22}$. In particular, \cref{fig:JohannsenLR} shows the plots of the light-ring radius as a function of $a/M$ for different values of these parameters. 
For $\alpha_{22}=0$, increasing $\alpha_{13}$ enlarges the light-ring radius, reducing the orbital frequency. Through the eikonal correspondence~\eqref{eq:OREikonal}, this translates into a decrease of the real QNM frequency, reproducing the trend observed in \cref{fig:JohannsenFirstCase}. Conversely, with $\alpha_{13}=0$, increasing $\alpha_{22}$ decreases the light-ring radius, yielding higher orbital frequencies and correspondingly larger real parts of the QNMs, consistent with the behavior reported in \cref{fig:JohannsenA2}.


\section{Discussion and outlook}
\label{sec:Conclusions}
In this work, we have investigated an alternative strategy to estimate deviations from GR encoded in the gravitational QNMs of BH spacetimes beyond Kerr. Owing to its analytical simplicity, the eikonal approximation is commonly employed to study QNMs in modified-Kerr geometries. We have, instead, taken a complementary approach and computed exactly the QNMs of a test scalar field on the given background and used their deviations from the Kerr predictions as a proxy for the corresponding gravitational corrections. 

Our results show that this approach provides a reliable and practical estimate of beyond GR effects. For Kerr-Newman and EsGB BHs, scalar QNMs reproduce the exact gravitational corrections at the level of tens of percent. Given that current and near-future ringdown measurements constrain these deviations at the percent level, such an accuracy is fully adequate for phenomenological studies (see also~\cite{Tang:2025qaq} for a similar conclusion about using the WKB approximation to compute QNMs). Moreover, the scalar-field strategy is typically comparable to--and in some cases more accurate than--the eikonal approximation when the latter is applied outside its strict regime of validity (namely, for small-$\ell$ modes). Very recently, the eikonal and post-Kerr~\cite{Glampedakis:2017dvb} approximations have been benchmarked against exact scalar QNMs, showing overall good agreement~\cite{DeSimone:2026mkz}.

Motivated by these findings, we have used test scalar QNMs to explore deviations from the Kerr metric in setups where a fundamental theoretical description is lacking. This is particularly relevant for phenomenologically deformed models, in which deviations are controlled by one or more parameters without direct reference to an underlying theory. Although such metrics do not constitute explicit tests of specific extensions of GR, they provide a valuable framework to characterize and constrain generic departures from Kerr in a model-independent way.

Given its relevance for tests of gravity based on BH imaging, we have focused on the Johannsen family of spacetimes. By isolating individual deformation parameters, we have shown that ringdown observables respond differently to them, in a manner that can be qualitatively interpreted within the eikonal framework. Our analysis enables a direct comparison between ringdown tests and BH shadow observations. We find that current ringdown constraints are already comparable to, and in some cases more stringent than, those derived from present EHT data. 
While we focused on a specific family of parametrized metrics, we expect these qualitative conclusions to hold more generally for other phenomenologically deformed spacetimes.

Furthermore, the accuracy of ringdown tests will improve significantly in the next years. Next-generation ground-based detectors such as the Einstein Telescope and Cosmic Explorer can reach ringdown signal-to-noise ratio $\approx100$ for several events per year~\cite{Bhagwat:2023jwv,ET:2025xjr}, whereas space missions like LISA can even achieve ringdown signal-to-noise ratio $\approx1000$ in certain optimistic scenarios~\cite{Berti:2016lat,Bhagwat:2021kwv,LISA:2024hlh}.
These observations will allow for ringdown tests at the subpercent level. Such an exquisite accuracy represents also a challenge for ringdown modelling~\cite{Crescimbeni:2025ytx} and the test scalar-field shortcut proposed here might become too inaccurate in that case.
It would be interesting to improve our method (e.g., using test spin-2 fields, see below) and compare future projected ringdown constraints with those expected from next-generation BH imaging facilities~\cite{Ayzenberg:2023hfw,Uniyal:2025uvc}.

Our approach also makes it clear that ringdown measurements can probe deviations of the spacetime geometry that remain invisible to BH imaging observations. GW tests and EHT measurements should therefore be regarded as complementary probes of the strong-field regime: taken together, they provide a more complete and robust assessment of possible departures from the Kerr paradigm. In this sense, our analysis can be viewed as extending the perspective of Ref.~\cite{Volkel:2020xlc}, which investigated how EHT observations can constrain deviations from the Kerr geometry parametrized through post-Newtonian expansions, focusing primarily on static metrics. While that work already pointed out the complementarity with GW observations, our results make this connection explicit by showing how ringdown measurements of spinning BHs can probe geometric deformations that are largely or totally inaccessible to BH imaging.

Finally, we have focused here on a test scalar field, but the method can be straightforwardly extended to test vector or tensor fields. Even when the Klein-Gordon equation separates in a given spinning geometry (as occurs, for example, in the Johannsen metric discussed above), separability of the equations for spin-1 and spin-2 perturbations is not guaranteed. Nevertheless, we expect that using a test spin-2 field as a proxy would yield more accurate results. Quantifying this expectation would be an interesting direction for future work.


\begin{acknowledgements}
We thank the Simons Center for Geometry and Physics for hospitality during the program \emph{50 years of the Black Hole Information Paradox}, where this project was initiated.
We thank Emanuele Berti for gently providing us with the data from Ref.~\cite{Berti:2005eb} and Mariafelicia de Laurentis for useful discussions. 
This work is supported by the MUR FIS2 Advanced Grant ET-NOW (CUP:~B53C25001080001) and by the INFN TEONGRAV initiative. We also gratefully acknowledge support from a research grant funded under the INFN–ASPAL agreement as part of the Einstein Telescope training program.
\end{acknowledgements}

\section*{Data Availability} 
The data generated to produce the plots in \cref{fig:JohannsenFirstCase,fig:JohannsenA2,fig:JohannsenA5,fig:Mixed} are publicly available at the link~\cite{Bitbucket_repo}.

\begin{appendix}

\section{Lightning review of the eikonal approximation}
\label{App:EikonalApproximation}

In this appendix we briefly review the eikonal approximation for the computation of BH QNMs~\cite{Press:1971wr,Ferrari:1984zz,Mashhoon:1985cya,Cardoso:2008bp} (see also Ref.~\cite{Yang:2012he} and Appendix A of Ref.~\cite{Cardoso:2016olt} for explicit calculations in the Kerr-Newman spacetime). This approach provides a geometric interpretation of the QNM spectrum in the large-angular-momentum limit and offers a computationally simple way to estimate the dominant modes.

The key ingredient of the approximation is the correspondence between QNMs in the large-$\ell$ limit and the motion of null particles at unstable circular null geodesics. Since in the main text we focus on modes with $\ell = |m|$, it is sufficient to restrict attention to the equatorial motion $\theta = \pi/2$.

For a general stationary, axisymmetric metric of the form
\begin{equation}
    \dd s^2 = g_{tt} \dd t^2 + 2 g_{t\phi} \dd t \dd \phi + g_{rr} \dd r^2 + g_{\theta\theta} \dd \theta^2 + g_{\phi\phi} \dd \phi^2
\end{equation}
the radial motion of null particles is governed by the equation $\dot r^2 + V_\text{eff} = 0$, where the dot denotes derivation with respect to an affine parameter, while $V_\text{eff}$ is the effective potential
\begin{equation}
    V_\text{eff}\equiv \frac{\mathcal{E}^2 g_{\phi\phi} + 2 \mathcal{E} \mathcal{L} g_{t\phi} + \mathcal{L}^2g_{tt}}{g_{rr} g^2_{t\phi} - g_{tt} g_{rr} g_{\phi\phi}}\, .
    \label{eq:AppVEff}
\end{equation}
This expression is understood to be evaluated at $\theta = \pi/2$. Here $\mathcal{E}$ and $\mathcal{L}$ denote the conserved energy and angular momentum of the particle, respectively.

The location of the circular photon orbits $r_\text{LR}$, together with $\mathcal{L}/\mathcal{E}$, are solutions of the equations
\begin{equation}
    V_\text{eff}(r_\text{LR}) = V_\text{eff}'(r_\text{LR}) = 0\, ,
    \label{eq:AppLR}
\end{equation}
where the prime denotes derivation with respect to $r$. In rotating spacetimes, \cref{eq:AppLR} admits two equatorial solutions, corresponding to null particles whose angular momentum is either aligned (prograde) or anti-aligned (retrograde) with the BH spin. Since in the present work we focus on modes with $m=\ell$ and positive BH spin, we restrict our attention to the prograde orbit. In the following, both in this appendix and in the main text, the term ``light ring'' will refer exclusively to this prograde orbit.

Within the eikonal approximation, the real part of the QNM frequency is related to the orbital frequency of null particles at $r = r_\text{LR}$,
\begin{equation}
    \Omega = \left.
    \frac{-g'_{t\phi} + \sqrt{g'^{2}_{t\phi}-g'_{tt} g'_{\phi\phi}}}{g'_{\phi\phi}} \right|_{\substack{\theta=\pi/2 \\  r=r_{\rm LR}}}\, .
\end{equation}
Then, the real part of the QNM frequency reads
\begin{equation}
    \omega_\text{R} \sim \ell \Omega \, .
    \label{eq:OREikonal}
\end{equation}
The imaginary part of the QNM frequency is instead determined by the Lyapunov exponent $\lambda$, which quantifies the instability timescale of the null orbit,
\begin{equation}
\begin{split}
    \lambda &= \left.-\frac{1}{\dot t}\sqrt{\frac{V''_\text{eff}}{2}} \right|_{\substack{\theta=\pi/2 \\  r=r_{\rm LR}}}\, ; \\
    \dot t &= -\frac{\mathcal{E}^2 g_{\phi\phi} + \mathcal{L}g_{t\phi}}{g^2_{t\phi}-g_{tt} g_{\phi\phi}}\, ,
\end{split}
\label{eq:AppImaginaryQNM}
\end{equation}
leading to
\begin{equation}
    \omega_\text{I} = -\left(n+\frac{1}{2} \right) |\lambda|\, .
    \label{eq:OIEikonal}
\end{equation}

\section{Comparison between scalar, gravitational and eikonal QNMs for a Kerr-Newman BH: results at different spins}
\label{App:KNComparisonAddition}

\begin{figure*}[h!]
    \centering
    \includegraphics[width=0.32\textwidth]{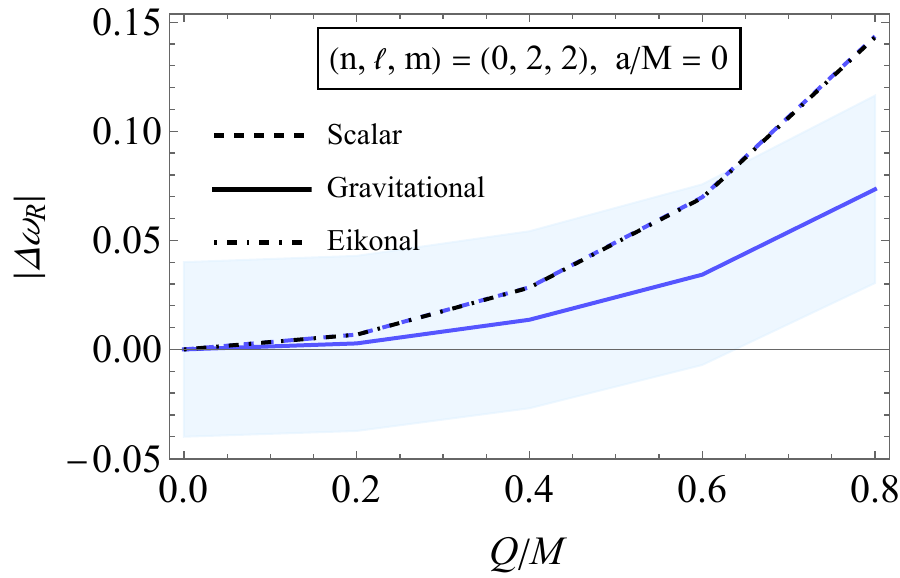}
    \includegraphics[width=0.32\textwidth]{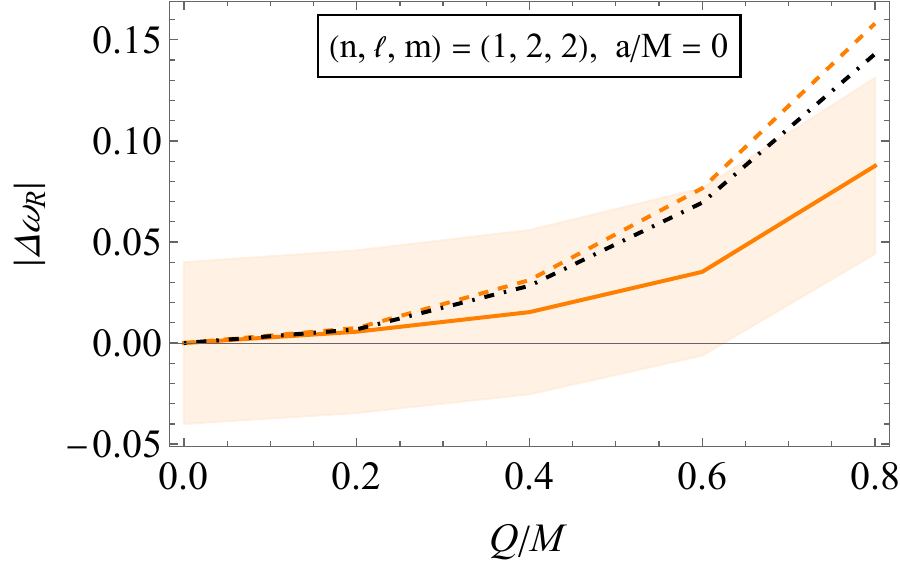}
    \includegraphics[width=0.32\textwidth]{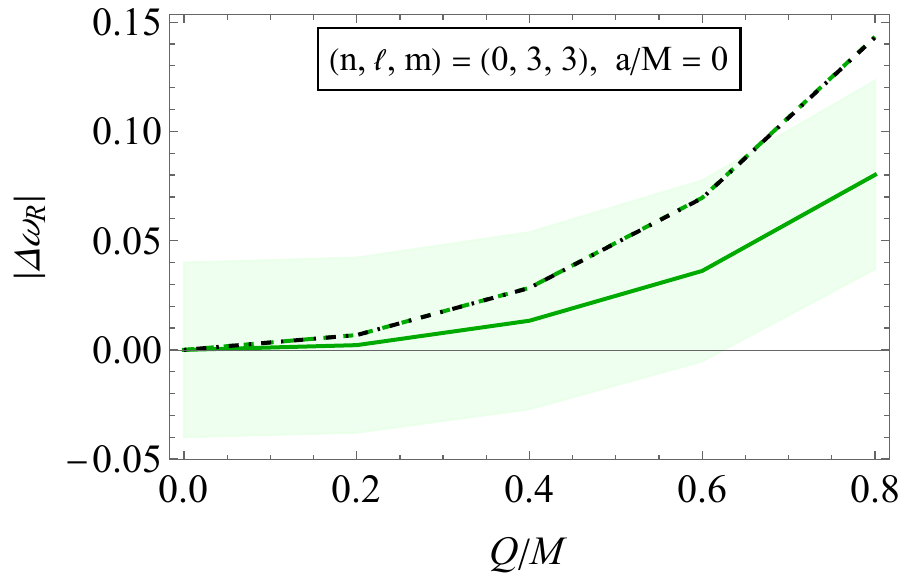}\\
    \includegraphics[width=0.32\textwidth]{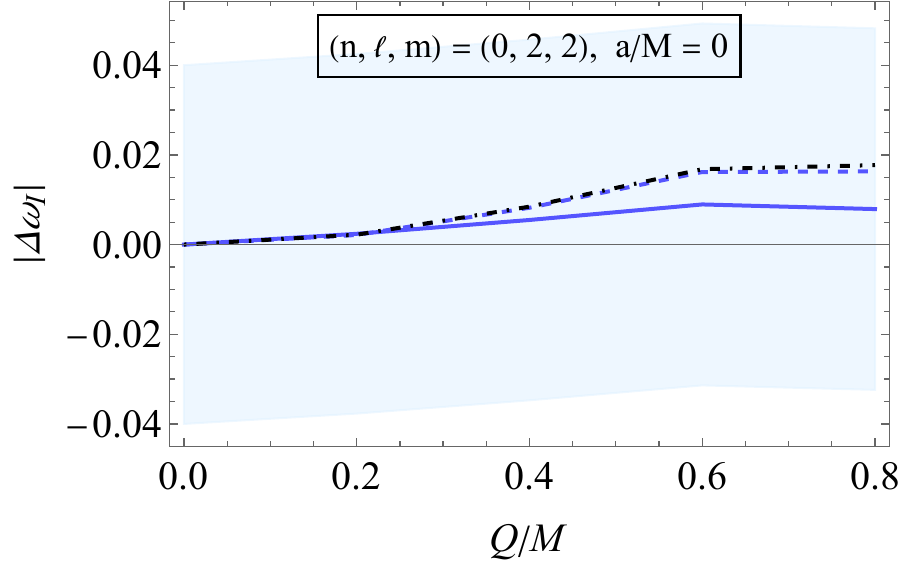}
    \includegraphics[width=0.32\textwidth]{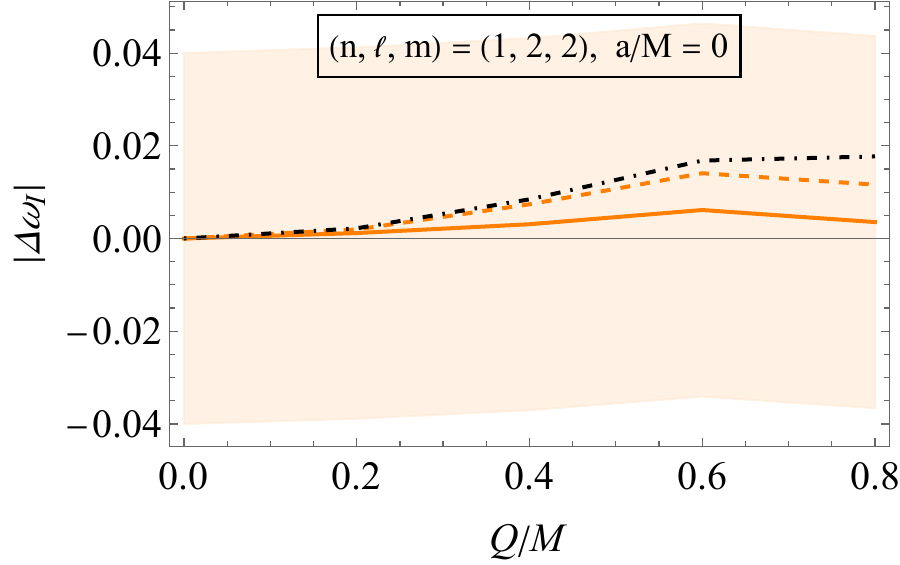}
    \includegraphics[width=0.32\textwidth]{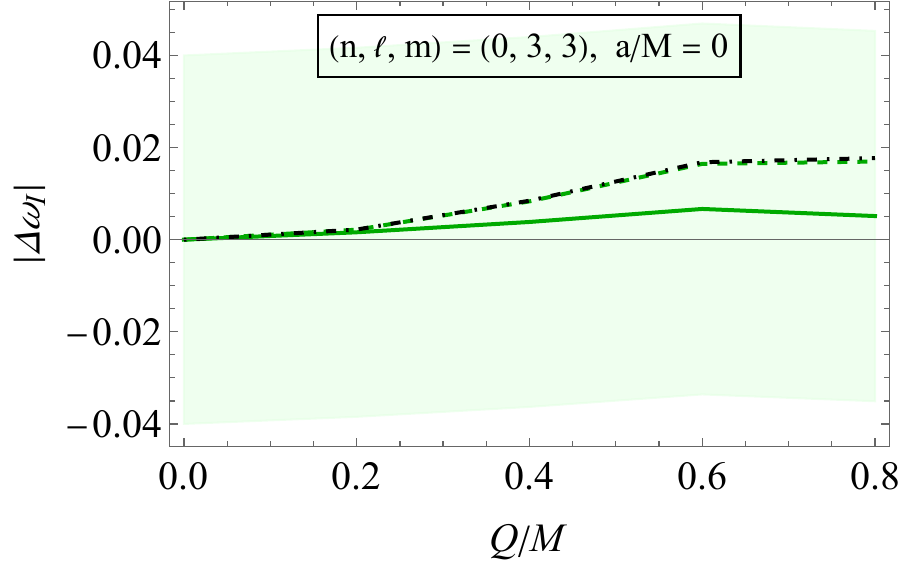}\\
    \caption{Plots of the absolute values of the relative deviations of the considered Kerr-Newman QNMs from Kerr results, shown as functions of $Q/M$. Results of both real and imaginary parts are shown. For all figures, we fixed $a/M = 0$. Solid and dashed lines refer to gravitational and scalar results, while the dot-dashed black line to eikonal ones. Shaded regions indicate bands around the gravitational results for the real and imaginary parts; their widths are given by the right-hand-side of \cref{eq:BandWidth} with $X = 4\, \%$.}
    \label{fig:KN_a0}
\end{figure*}

\begin{figure*}[h!]
    \centering
    \includegraphics[width=0.32\textwidth]{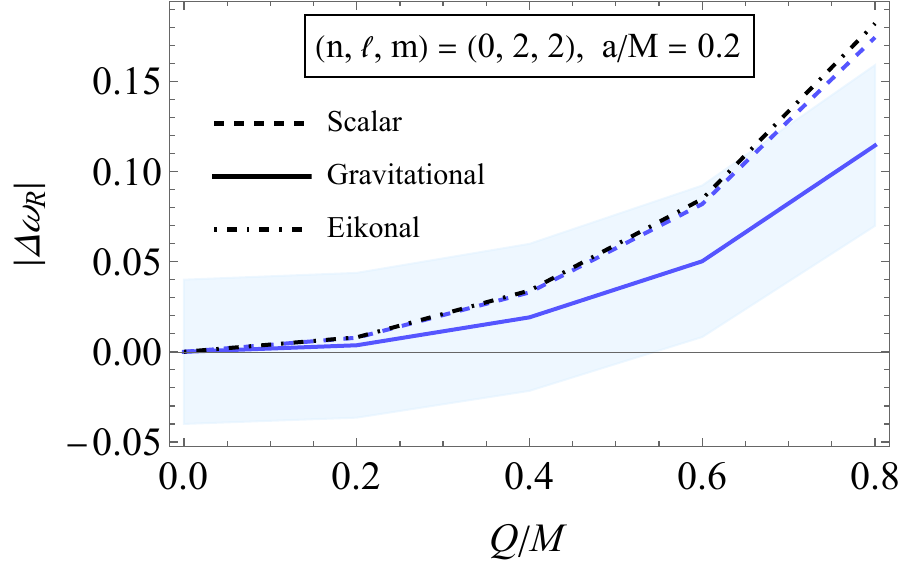}
    \includegraphics[width=0.32\textwidth]{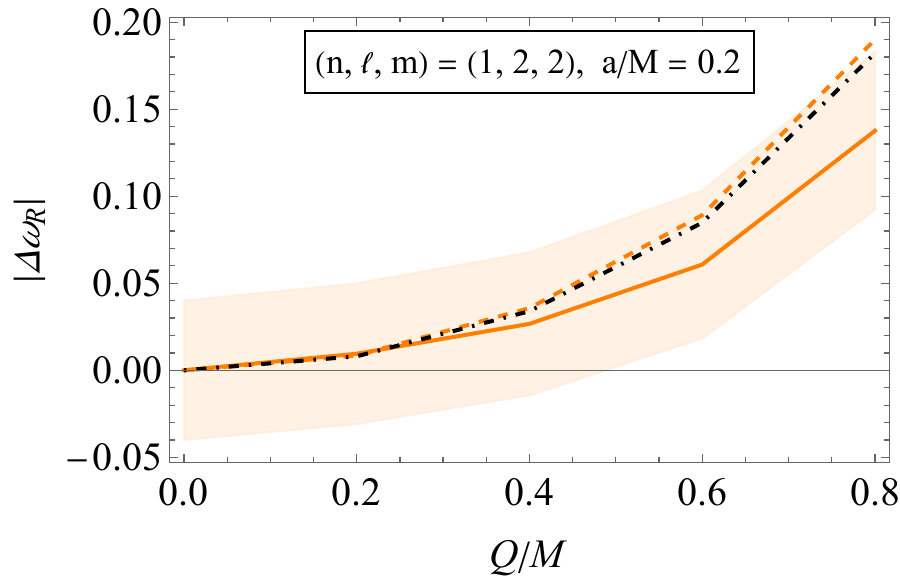}
    \includegraphics[width=0.32\textwidth]{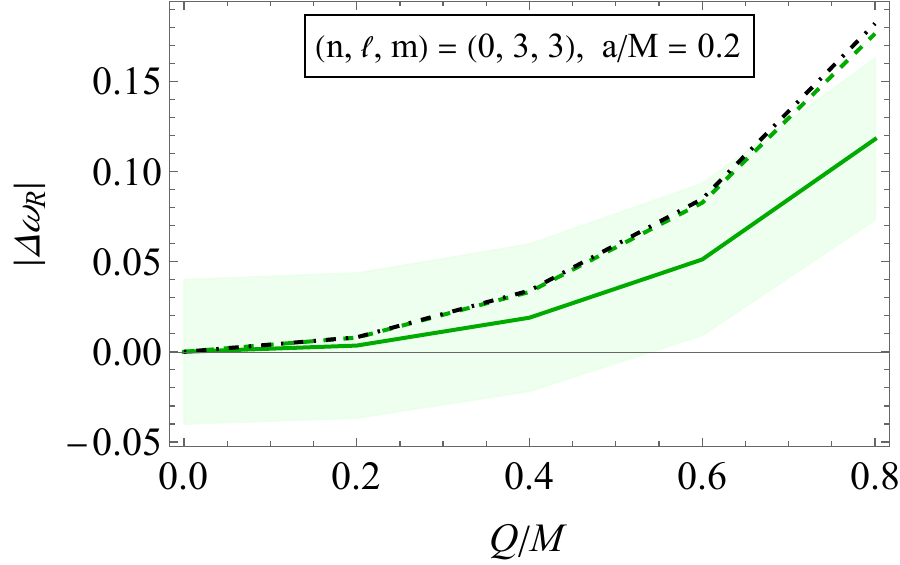}\\
    \includegraphics[width=0.32\textwidth]{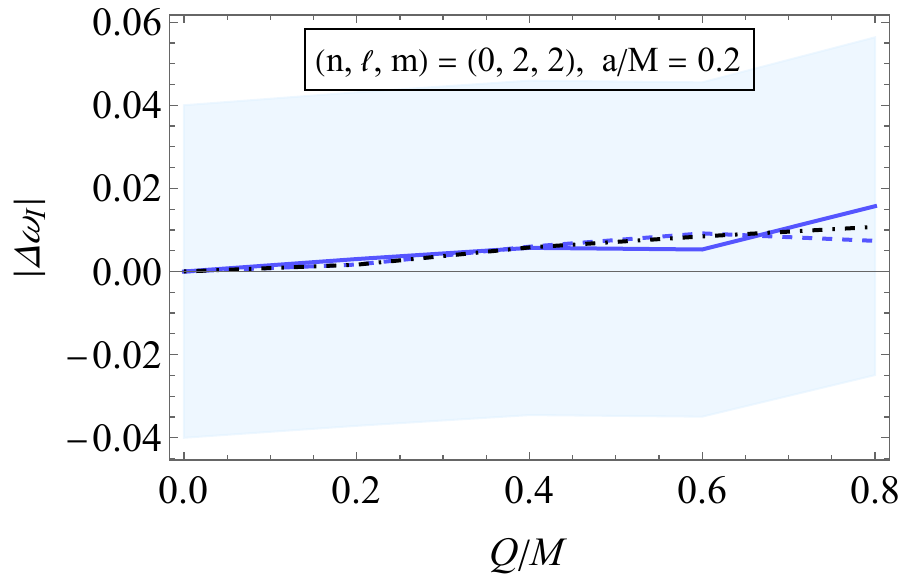}
    \includegraphics[width=0.32\textwidth]{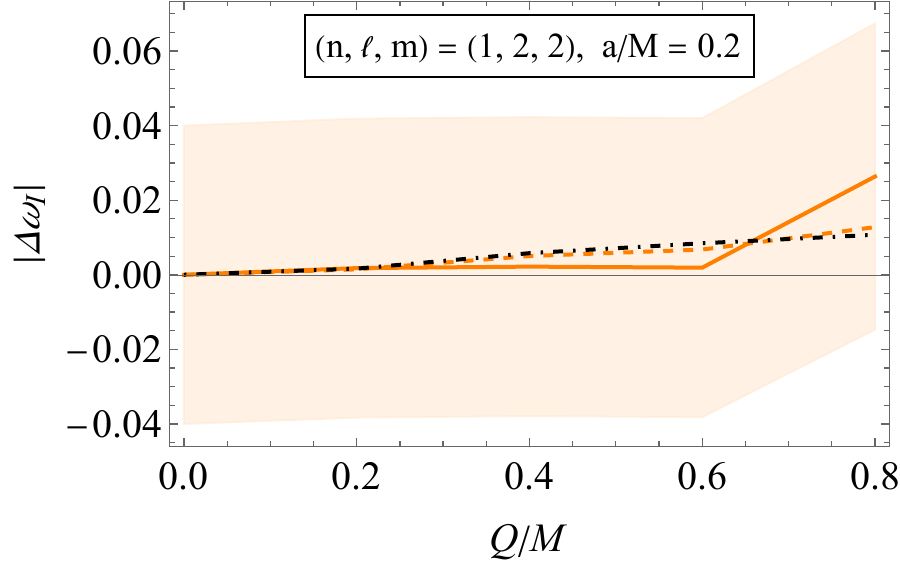}
    \includegraphics[width=0.32\textwidth]{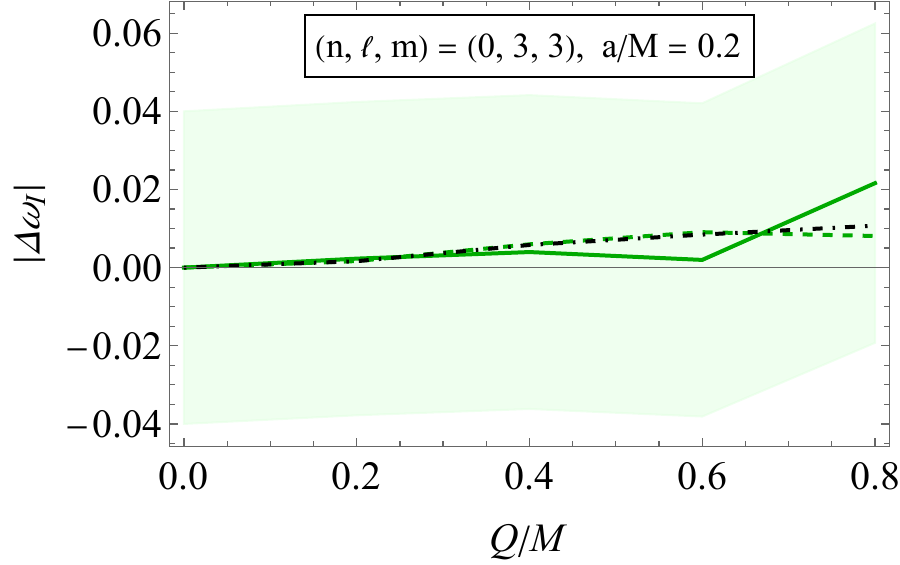}\\
    \caption{Plots of the absolute values of the relative deviations of the considered Kerr-Newman QNMs from Kerr results, shown as functions of $Q/M$. Results of both real and imaginary parts are shown. For all figures, we fixed $a/M = 0.2$. Solid and dashed lines refer to gravitational and scalar results, while the dot-dashed black line to eikonal ones. Shaded regions indicate bands around the gravitational results for the real and imaginary parts; their widths are given by the right-hand-side of \cref{eq:BandWidth} with $X = 4\, \%$.}
    \label{fig:KN_a02}
\end{figure*}

Here we present complementary results to those discussed in \cref{Sec:KerrNewman_ScalarvsGravModes} by considering two additional values of the dimensionless spin, namely $a/M = 0$ and $a/M = 0.2$. The corresponding results are shown in \cref{fig:KN_a0} for the nonrotating case and in \cref{fig:KN_a02} for $a/M = 0.2$, respectively.
Also for these values of $a/M$, we notice similar results as those discussed in \cref{Sec:KerrNewman_ScalarvsGravModes}, with both the eikonal and scalar predictions on deviations from Kerr being consistent within the bands with gravitational results.

\section{Detailed calculations of the modified scalar Teukolsky equations in the Johannsen spacetime}
\label{App:KGJohannsen}

We start from the Klein-Gordon equation \eqref{eq:KGequation}, which, after some calculations, can be written as
\begin{equation}
\begin{split}
    &\sqrt{A_5} \partial_r \left[\sqrt{A_5} \frac{\tilde \Sigma}{B(r, \theta)} \Delta \partial_r \Psi \right] + \\
    &\frac{1}{\sin\theta}\partial_\theta \left[\sin \theta \frac{\tilde \Sigma}{B(r, \theta)} \partial_\theta \Psi \right] \\
    &+ \frac{a^2 \Delta \sin^2 \theta - A_1^2 (r^2+a^2)^2}{\Delta}\frac{\tilde \Sigma}{B(r, \theta)} \partial^2_t \Psi\\ 
    &-2\frac{a\left[(r^2+a^2)A_1 A_2 - \Delta \right]}{\Delta} \frac{\tilde \Sigma}{B(r, \theta)} \partial_t \partial_\phi \Psi \\
    &+ \frac{\Delta - a^2 A_2^2 \sin^2 \theta}{\Delta}\frac{\tilde \Sigma}{B(r, \theta)} \frac{\partial^2_\phi \Psi}{\sin^2 \theta} = 0\, ,
\end{split}
\label{eq:KGFirstStep}
\end{equation}
where we have defined
\begin{equation}
    B(r, \theta) \equiv (r^2 + a^2)A_1 - a^2 A_2 \sin^2 \theta\, .
\end{equation}
Equation~\eqref{eq:KGFirstStep} contains terms whose coefficients depend purely on $r$ or $\theta$, as well as terms mixing the two variables, which in general obstruct separability. The obstruction originates from the factor $\tilde\Sigma/B(r,\theta)$. Therefore, separability can be enforced by assuming that this quantity factorizes as
\begin{equation}
    \frac{\tilde \Sigma}{B} \equiv \tilde R(r) \tilde \Theta(\theta)\, .
    \label{eq:DefinitionsSeparation}
\end{equation}
One can verify explicitly that, under this assumption, all $r$ and $\theta$ dependencies in \eqref{eq:KGFirstStep} can be consistently separated. 

To determine the explicit form of $\tilde R$ and $\tilde \Theta$, we use \eqref{eq:DefinitionsSeparation} together with the definitions of $B(r,\theta)$ \eqref{eq:DefinitionsSeparation}, and $\tilde \Sigma$ in \cref{eq:ModificationsKerrJohannsen}. This leads to
\begin{equation}
\begin{split}
    r^2 + f(r) + a^2 \cos^2 \theta = &\tilde R \tilde \Theta \left[(r^2+a^2)A_1- a^2 A_2 \sin^2 \theta\right]\, .
    \label{eq:DefinitionsSeparation2}
\end{split}
\end{equation}
Since the left-hand side does not contain mixed $r$-$\theta$ terms, consistency requires $\tilde \Theta(\theta) = 1$ and $\tilde R = A_2(r)^{-1}$. With these identifications, \cref{eq:DefinitionsSeparation2} yields the condition~\eqref{eq:ConstraintfrSeparability}, which coincides with the result obtained in Ref.~\cite{Konoplya:2018arm}. 

Substituting \cref{eq:ConstraintfrSeparability} back into \cref{eq:KGFirstStep}, and multiplying the resulting equation by $A_2$, we obtain
\begin{equation}
\begin{split}
    &A_2 \sqrt{A_5} \partial_r\left[\frac{\sqrt{A_5}}{A_2} \Delta \, \partial_r \Psi \right] + \frac{1}{\sin\theta} \partial_\theta \left(\sin \theta \partial_\theta \Psi \right) \\
    &+ a^2 \sin^2 \theta \partial^2_t \Psi - \frac{(r^2+a^2)^2}{\Delta}A_1 \partial^2_t \Psi \\
    &-2\frac{a\left[(r^2+a^2)A_1 A_2 - \Delta \right]}{\Delta} \partial_t \partial_\phi \Psi + \frac{\partial^2_\phi \Psi}{\sin^2 \theta} \\
    &-\frac{a^2 A_2^2}{\Delta} \partial^2_\phi\Psi = 0\, .
\end{split}
\end{equation}
Upon inserting the separable ansatz \eqref{eq:AnsatzPsi}, dividing by $R(r)S(\theta)$, and simplifying common factors, we arrive at
\begin{equation*}
\begin{split}
    &\frac{A_2 \sqrt{A_5}}{R}\partial_r \left(\frac{\sqrt{A_5}}{A_2} \, \Delta \, R' \right) + \frac{1}{S} \frac{1}{\sin\theta}\partial_\theta \left(\sin \theta \partial_\theta S \right) \\
    &- a^2 \omega^2 + a^2 \omega^2 \cos^2 \theta +\frac{(r^2+a^2)^2\omega^2}{\Delta} A_1\\
    &-\frac{2am \omega}{\Delta}\left[(r^2+a^2)A_1 A_2 - \Delta \right]-\frac{m^2}{\sin^2 \theta} + \frac{a^2 m^2}{\Delta}A_2^2 = 0\, .
\end{split}
\end{equation*}
Each term now depends either on $r$ or on $\theta$ only, allowing the equation to be written schematically as
\begin{equation}
    \mathcal{R}(r) + \mathcal{T}(\theta) = 0\, .
\end{equation}
Consistency then requires
\begin{equation}
    \mathcal{R}(r) = \Lambda_{\ell m}\, , \qquad \mathcal{T}(\theta) =-\Lambda_{\ell m}\, ,
\end{equation}
where $\Lambda_{\ell m}$ is the separation constant. This procedure leads directly to the angular and radial equations \eqref{eq:Teu1} and \eqref{eq:Teu2}.

\section{QNMs in the Johannsen metric: a multi-parameter example}
\label{app:MixedConstantsJohannses}

In the following, we investigate the impact on the QNM frequencies of simultaneously varying multiple deformation parameters. Since our aim is to compare the effects of these parameters on the ringdown and on the shadow, and since we have shown that the parameter $\alpha_{52}$ does not affect the location of the light ring and has a negligible impact on the real part of the QNM frequencies of the Johannsen metric (see \cref{subsubsec:A5}), we restrict our analysis to the simultaneous variation of $\alpha_{13}$ and $\alpha_{22}$.

\begin{figure}
    \centering
    \includegraphics[width=\linewidth]{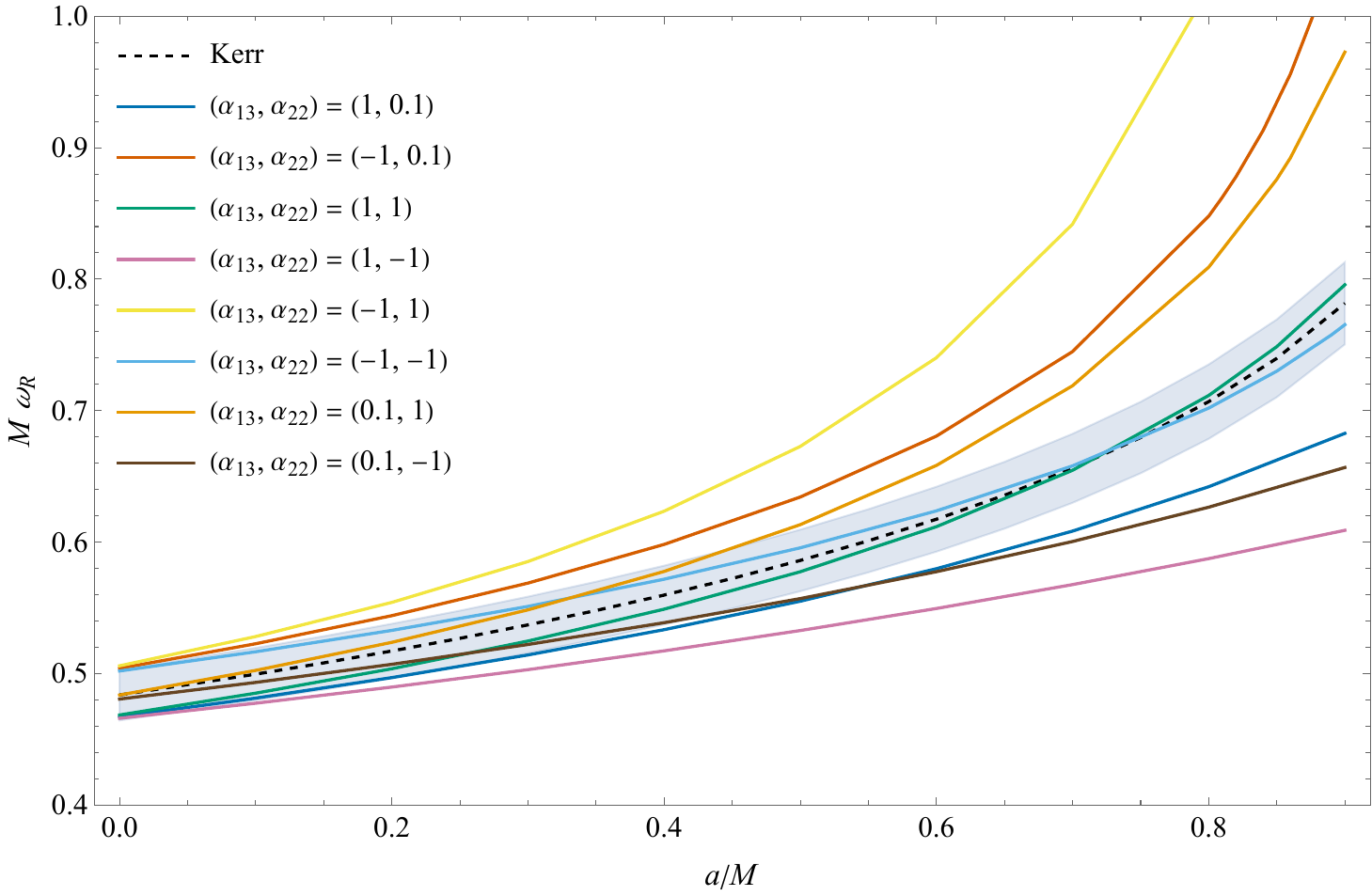}\hfill
    \includegraphics[width=\linewidth]{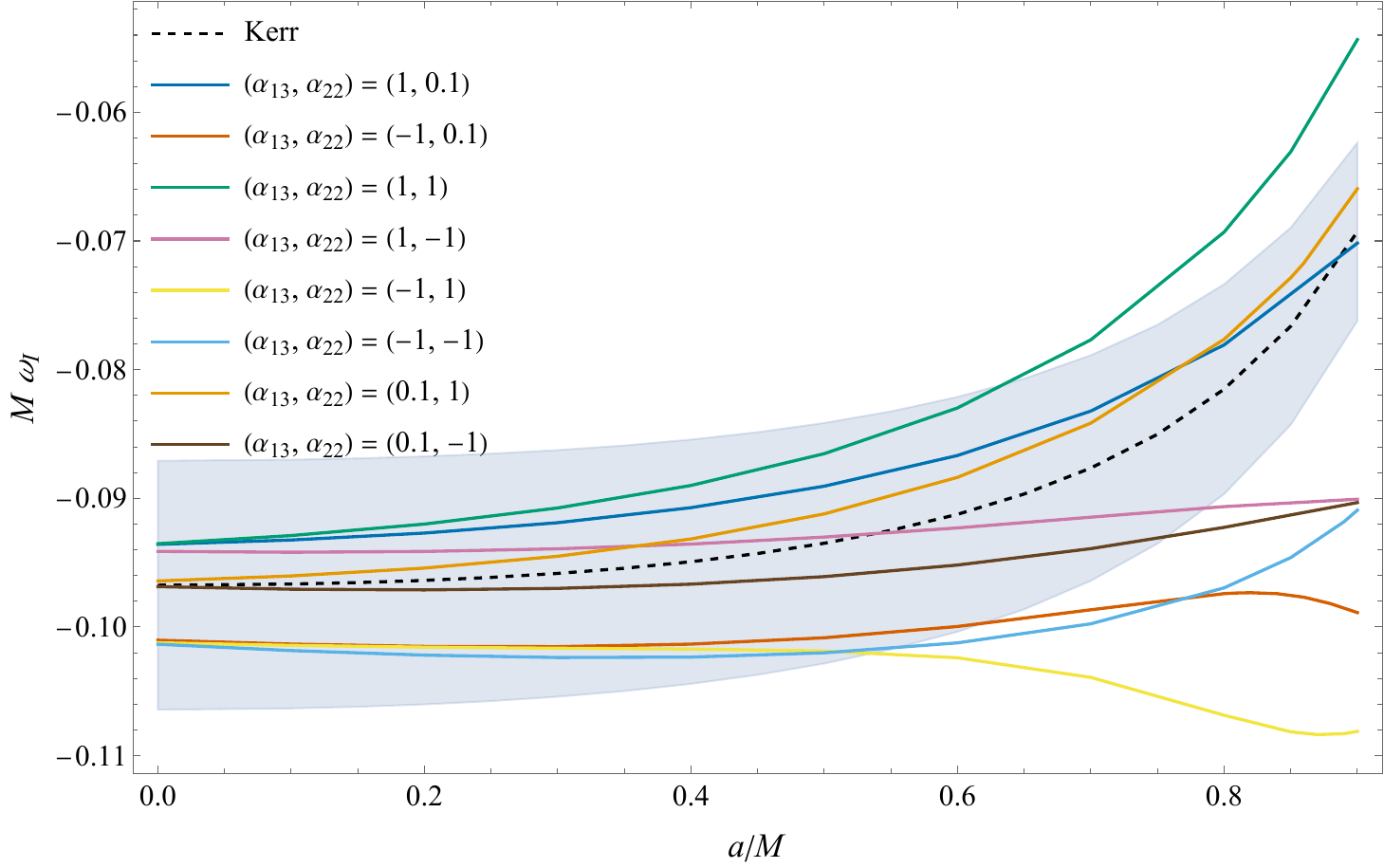}
        \caption{Real (top) and imaginary (bottom) parts of the QNM frequencies for the Johannsen metric \eqref{eq:JohannsenMetric} with $A_5=1$. The frequencies are shown as functions of the dimensionless spin $a/M$ for different combinations of the deformation parameters $\alpha_{13}$ and $\alpha_{22}$. The corresponding Kerr scalar QNM frequencies are shown for comparison (black dashed line). The blue shaded regions indicate the $\pm 4\%$ and $\pm 10\%$ bands around the Kerr values for the real and imaginary parts, respectively.}
    \label{fig:Mixed}
\end{figure}

As representative cases, we consider combinations of positive and negative values of the deformation parameters, with magnitudes of order $\mathcal{O}(10^{-1})$ and $\mathcal{O}(1)$, consistent with the bounds discussed in \cref{subsubsec:TheoreticalConstraints} and \cref{sec:ObservationalConstraints}. The corresponding QNM frequencies are computed using the numerical strategy described in \cref{subsubsec:SolutionStrategy}. The resulting spectra are reported in \cref{fig:Mixed}.
Overall, the resulting behavior is consistent with and complements the findings of \cref{subsubsec:A1} and \cref{subsubsec:A2}. In the main text, we showed that varying $\alpha_{13}$ and $\alpha_{22}$ individually produces different trends in the QNM frequencies, depending on the sign of the corresponding deformation parameter. When both parameters are varied simultaneously, the resulting curves display a superposition of the features observed in the single-parameter cases. These effects are, as expected, negligible when one parameter is one order of magnitude smaller than the other, while more pronounced combined effects emerge when both parameters are of $\mathcal{O}(1)$.

The combined effects are particularly evident for the real part of the frequencies. When $\alpha_{13}$ and $\alpha_{22}$ have the same sign and are both of order $\mathcal{O}(1)$, the corresponding curves remain close to the Kerr prediction. This behavior is consistent with the opposite trends observed in the single-parameter analysis. In particular, as discussed in \cref{subsubsec:A1}, increasing positive values of $\alpha_{13}$ shift the curves towards smaller frequencies, while increasing negative values shift them towards larger frequencies. Conversely, when only $\alpha_{22}$ is non-zero, the behavior is reversed: increasingly positive (negative) values of $\alpha_{22}$ shift the curves towards larger (smaller) frequencies. Therefore, when $\alpha_{13}$ and $\alpha_{22}$ are varied simultaneously with the same sign, their individual effects partially cancel each other, leading to a reduced deviation from the Kerr results.

When the two deformation parameters have opposite signs, instead, their contributions generally reinforce each other. In particular, when both parameters are of order $\mathcal{O}(1)$, the combined effect can drive the curves outside the adopted tolerance bands, excluding a larger region of the parameter space compared to the cases where only one deformation parameter is non-zero.

Analogous combined effects are observed for the imaginary part of the frequencies, although with a different trend. In this case, deformation parameters with the same sign produce larger deviations than those obtained by varying the parameters individually, whereas a partial cancellation occurs for opposite relative signs. Overall, unlike the real part of the frequencies, the net effect on the imaginary part does not lead to substantial differences with respect to the single-parameter analyses presented in \cref{subsubsec:A1} and \cref{subsubsec:A2}.

\end{appendix}

 \newpage

\bibliographystyle{apsrev4-1}
\bibliography{Refs.bib}

\end{document}